\pdfoutput=1
\documentclass[11pt]{article}

\usepackage{enumitem}
\usepackage{microtype}
\usepackage[tt=false]{libertine}
\usepackage{tabularx}
\usepackage{amssymb}

\usepackage{tikz}
\usetikzlibrary{arrows.meta}

\usepackage{caption}
\usepackage{graphicx}
\usepackage{subcaption}

\newcommand{\st}[1]{\textsf{#1}}

\usepackage{graphicx}
\usepackage[breaklinks=true, linktocpage=true,]{hyperref}
\usepackage{amssymb}
\usepackage{tikz}
\usepackage{commath}
\usepackage{braket}
\usepackage{wasysym}
\usepackage[margin=1in]{geometry}
\usepackage{booktabs}
\usepackage{amsmath}
\usepackage{amsthm}
\usepackage{lipsum}
\usepackage{url}
\usepackage[boxed]{algorithm2e}
\usepackage{algpseudocode}

\usepackage{aliascnt}
\usepackage[nameinlink,noabbrev]{cleveref}

\newcommand{\alg}{\mathsf{ALG}}
\newcommand{\roc}{\calX \rightarrow \zo^\lambda}
\newcommand{\ideal}{\calR_{\textsc{ideal}}}

\usepackage[
	lambda,
	operators,
	advantage,
	sets,
	adversary,
	landau,
	probability,
	notions,
	logic,
	ff,
	mm,
	primitives,
	events,
	complexity,
	asymptotics,
	keys]{cryptocode}

\usepackage{bbm}
\usepackage{todonotes}

\addtocontents{toc}{\protect\setcounter{tocdepth}{2}}

\renewcommand{\epsilon}{\varepsilon}

\newcommand{\R}{\mathbb{R}}

\newcommand{\E}{\mathop{{}\mathbb{E}}}

\DeclareMathOperator{\calP}{\mathcal{P}}

\DeclareMathOperator{\calU}{\mathcal{U}}

\DeclareMathOperator{\calX}{\mathcal{X}}
\DeclareMathOperator{\calR}{\mathcal{R}}

\DeclareMathOperator{\calT}{\mathcal{T}}

\renewcommand{\Pr}{\operatorname{{\bf Pr}}}

\newcommand{\Com}{\mathsf{Commit}}
\newcommand{\Ver}{\mathsf{Verify}}
\newcommand{\Acc}{\mathsf{Acc}}

\newcommand{\Bad}{\mathsf{Bad}}
\DeclareMathOperator{\dom}{dom}

\newcommand{\Norm}{\textup{\textsc{Harden}}}

\newtheorem{theorem}{Theorem}
\newtheorem{corollary}{Corollary}

\newtheorem{lemma}{Lemma}[section]

\newtheorem{remark}{Remark}

\newtheorem{definition}{Definition}

\newtheorem{fact}{Fact}

\Crefname{fact}{Fact}{Facts}

\newcommand{\zo}{\{0,1\}}

\title{Merkle Trees Minimize Verifier Complexity in the ROM}

\author{
\shortstack{
  Miranda Christ\\
  MIT and RESI\\
  {\small \texttt{mchrist@cs.columbia.edu}}
}
\and
\shortstack{
  Oliver Korten\\
  Institute for Advanced Study\\
  {\small \texttt{korten@ias.edu}}
}
\and
\shortstack{
  Tal Malkin\\
  Columbia University\\
  {\small \texttt{tal@cs.columbia.edu}}
}
\and
\shortstack{
  Noam Mazor\\
  New York University\\
  {\small \texttt{noammaz@gmail.com}}
}
\and
\shortstack{
  Shyamal Patel\\
  UT Austin\\
  {\small \texttt{shyamalpatelb@gmail.com}}
}
}
\date{}

\begin{document}
\maketitle

\begin{abstract}
    Nearly five decades after their introduction, Merkle trees remain the most efficient hash-based vector commitments.
    They compress a vector $x \in \zo^n$ into a commitment whose size is independent of $n$, such that a verifier can open any bit $x_i$ using $\log n$ hash evaluations. The security of the scheme (intractability of finding two distinct openings for a given bit using the same commitment) follows from the collision resistance of the hash function.
    The verifier's runtime complexity is a bottleneck in many key applications of vector commitments, including SNARKs, credentials, and authenticated dictionaries.
    As such, a longstanding open question has been whether it is possible to construct hash-based vector commitments where the verifier makes a sublogarithmic number of hash evaluations.

    We prove that sublogarithmic verifier complexity is impossible to achieve through black-box use of a hash function; we not only rule out constructions from collision resistant hash functions, but in fact show that commitment schemes with sublogarithmic verifier complexity are unachievable in the random oracle model. In particular, any vector commitment with deterministic verifiers has verifier query complexity $\Omega_{\theta}(\log n)$, where $\theta$ is the input length of the random oracle.
    This bound is matched up to constant factors by high-arity Merkle trees.
    The key technical lemma in our proof is a \emph{disjoint product lemma}, which can be viewed as a computational analog of the van den Berg-Kesten-Reimer (BKR) inequality.
    We believe this lemma is of independent interest, and in particular may be useful for showing other lower bounds and separations in the random oracle model.
\end{abstract}

\section{Introduction}
The Merkle tree~\cite{merkle1987digital} is arguably the most ubiquitous cryptographic data structure. Merkle trees are widely used in practice, in Git,\footnote{Git objects documentation: \url{{https://git-scm.com/book/en/v2/Git-Internals-Git-Objects}}}
Google wallet,\footnote{Google wallet: \url{https://developers.google.com/wallet/identity/verify/accepting-ids-from-wallet-online}}
Bitcoin,\footnote{Bitcoin documentation: \url{{https://developer.bitcoin.org/reference/block_chain.html\#merkle-trees}}}
and internet browsers for certificate transparency.\footnote{Chrome certificate transparency: \url{https://googlechrome.github.io/CertificateTransparency/ct_policy.html}}
They are also a fundamental building block used in a variety of cryptographic protocols, including succinct non-interactive arguments (SNARKs), memory checkers, anonymous credentials, and registration-based encryption.
In these applications, Merkle trees provide the functionality of a \emph{vector commitment (VC)} or a \emph{cryptographic accumulator}.\footnote{For the sake of exposition, we focus on vector commitments. Our lower bound also applies to the weaker object of a cryptographic accumulator, which commits to a set and provides a membership witness for each element. Accumulators lack the ability of vector commitments to bind elements to indices, making them weaker and our lower bound stronger.}

A vector commitment produces a \emph{commitment}
$c$ to a given vector $x \in \{0,1\}^n$ and $n$ \emph{witnesses} certifying $x_i$ for each component $i \in [n]$.
A verifier, given a witness, can check whether a claimed index $i$ and value $x_i$ are consistent with the commitment.
Position binding ensures that forging a witness for the wrong entry $x_i' \neq x_i$ with respect to $c$ is computationally infeasible.

\begin{figure}[t]
  \centering
  \includegraphics[width=0.6\textwidth]{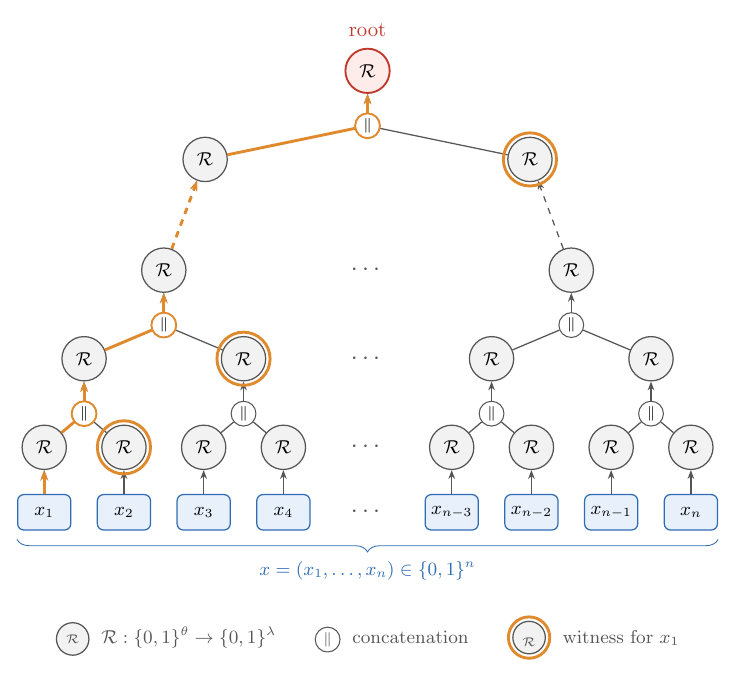}
  \caption{A binary Merkle tree over $x\in\{0,1\}^n$. Each internal node
    applies $\mathcal R$ to the concatenation of its two children.
    Highlighted in orange is the witness of $x_1$, consisting of the
    siblings along its path to the root.}
  \label{fig:merkle-binary}
\end{figure}

A Merkle tree uses a hash function $\calR: \zo^\theta \to \zo^\secpar$, often modeled as a \emph{random oracle} or truly random function.
The Merkle tree places the $n$ components at leaves, and each internal layer is formed by concatenating pairs in the previous layer and hashing them.
The eventual root of the tree is the commitment, and the witness of $x_i$ consists of the sibling nodes along its path to the root (shown in \Cref{fig:merkle-binary}).
The verifier checks this witness by recomputing these hashes and checking that each matches its parent; breaking position binding requires finding a collision.
In particular, any attacker making a polynomial number of queries has negligible probability of violating position binding, even if it uses unbounded computation alongside these queries.\footnote{Attack runtime and success probability are measured in terms of $\secpar$, the security parameter. We assume that $\theta$ and $n$ are larger than but polynomially bounded in $\secpar$.}

In applications, $n$ is typically quite large: the vector is, for example, a PCP or the set of all internet certificates.
As the verifier is often a weak client, the commitment size and the verifier runtime should be far smaller than $O(n)$.
In the random oracle model, this runtime is measured in the number of oracle queries the verifier makes.
For Merkle trees, the commitment size is $O(\secpar)$, and the verifier makes $\log n$ queries (the length of the leaf-to-root path).
Improving this verifier complexity to be sublogarithmic is a longstanding open question.

Interestingly, verifier complexity \emph{independent} of $n$ is possible if one is willing to use algebraically structured assumptions.
In particular, $O(\secpar)$-size commitments with $O(\secpar)$-time verifiers\footnote{Translating to time complexity, Merkle trees require $O(\secpar \log n)$-time verifiers; these algebraic constructions' lack of dependence on $n$ is far more desirable.} are enabled by RSA-based and pairing-based constructions~\cite{benaloh1993one,nguyen}.
Nonetheless, hash-based constructions are favored in practice as they offer plausible post-quantum security, avoid trusted setup (e.g., sampling an RSA modulus), and can in fact be faster: benchmarks~\cite{leung2022aardvark} show that Merkle verification is $400\times$ faster than Libert-Yung's~\cite{DBLP:conf/tcc/LibertY10} $O(1)$-time verification for 1,024-entry vectors.
Despite what the asymptotics suggest, $\log(n)$ hash evaluations are often cheaper in practice than a constant number of exponentiations and pairings.

Motivated by these practical advantages, a longstanding open question has been constructing more verifier-efficient vector commitments using only hash functions.
However, better schemes have remained elusive even in the \emph{random oracle model}, where the hash is modeled as a random function $\calR: \zo^\theta \rightarrow \zo^\secpar$ that serves as a perfect hash function (and in particular is easily shown to be collision resistant).
We show that this efficiency barrier is fundamental:

\begin{center}
    \emph{Merkle trees minimize verifier complexity in the random oracle model.}
\end{center}

That is, we show a lower bound on the number of verifier queries for \emph{any} vector commitment or accumulator with a deterministic verifier in the random oracle model.
This serves as a lower bound on constructions based on collision-resistant hash functions as well.
Specifically, we show:
    \begin{quote}
        Let $\theta$ be the random oracle input length. For any commitment size $\abs{c} = O(n^{\epsilon})$ for some constant $\epsilon < 1$, for sufficiently large $n$ relative to $\secpar$, the number of verifier queries per witness must satisfy
        \[
        k = \Omega(\log_{\theta} n).
        \]
    \end{quote}

This bound is matched up to constant factors by a high-arity Merkle tree. One can replace the binary tree with a tree of arity $\theta/\log^2 \theta$ by truncating the random oracle outputs to $\log^2 \theta$ bits, allowing us to pack $\theta/\log^2 \theta$ concatenated children into each query (see \Cref{fig:merkle-dary}).
Therefore, such high-arity trees achieve verifier query complexity $\Theta(\log_\theta n)$.

Note that although $\Omega(\log_\theta n)$ is constant if $n$ is upper bounded by a fixed polynomial, the constant depends on the degree of that polynomial.
Therefore, there is no constant $k$ that uniformly bounds the verifier complexity for all vector lengths.
For example, for $\theta = 2\secpar$, $n = \secpar^{d}$, and commitment size $O(\secpar)$, the verifier must make $d-1$ queries.

\subsection{Our techniques}
Our proof is surprisingly simple, combining two ingredients: a \emph{compression argument} and a new technical \emph{disjoint product lemma}.
We give a brief overview here and explain further in the technical overview (\Cref{sec:techo}).

Compression arguments are ubiquitous throughout combinatorics and TCS; they are used to give simplified ``operational'' descriptions of otherwise complicated probabilistic arguments. In the context of black-box separations in cryptography, they were first used in influential work of Gennaro and Trevisan~\cite{GT00}.
The idea is to use the object whose complexity we wish to lower bound to construct a communication protocol between Alice and Bob.
If this object is too efficient, the protocol allows Alice to send fewer bits than is information theoretically possible.

In our case, Alice wishes to communicate a uniformly random vector $x \in \zo^n$ to Bob, using a vector commitment whose verifier makes $k$ random oracle queries per witness.
Alice sends Bob a commitment to $x$ along with a \emph{small set of queries} from which Bob can derive \emph{all} of the verifier's queries.
Bob tests the verifier on each possible $(i, b) \in [n] \times \zo$ using the queries he derives.
Position binding ensures that he recovers $x$ based on the tuples the verifier accepts.
If Alice sends fewer than $n$ bits in total, we arrive at our information theoretic contradiction.

However, a priori, there is no reason why we should expect that a small set of queries allows Bob to derive all verifier queries---there are up to $kn$ total queries across the $n$ indices, far greater than $n/(\theta \cdot \text{query length})$.
The bad case for our communication protocol is where the witnesses' sets of queries have low overlap and it is difficult for Bob to guess these query sets himself.

This is where our second ingredient, the disjoint product lemma, enters. In the most extreme bad case, the queries needed to ``open'' all the bits $x_i$ successfully are pairwise disjoint, and hence sending them all would cost $kn\theta$ bits. However, recall that Alice needs to somehow construct a commitment which would allow Bob to open any individual bit successfully; in order to do this, she must essentially find these $n$ disjoint witnesses at the start. We will argue, using our disjoint product lemma, that if it is difficult for Bob to find these disjoint witnesses individually himself, then it must be \emph{extremely difficult} for Alice to find all of these disjoint witnesses simultaneously in the first place. Crucially, the hardness obtained from our lemma (where ``hardness'' is quantified by the probability that a bounded query procedure finds many disjoint witnesses) will be exponential in the number of disjoint witnesses found, which will allow us to union bound over Alice's effective advice. We hence conclude that this bad case occurs with negligible probability. Of course, this extreme case does not exhaust all possible situations in which the total number of oracle queries needed to open all bits exceeds $n$; our final argument will involve a more extensive case analysis, but this extreme case captures much of the difficulty of the problem.

We give a more detailed explanation of the lemma's application in \Cref{sec:techo} but state the lemma below:

\begin{lemma}[Disjoint Product Lemma]
\label{lem:crab-intro}
    Let $\calX$ be a set, $\calR: \roc$ be a random oracle, and $V^{\calR}$ be a deterministic verifier, which takes as input at most $k$ points $q_1, \dots, q_{k'} \in \calX$, queries $\calR(q_1), \dots, \calR(q_{k'})$ and then accepts or rejects. Let $t$ be an integer and suppose that any $(Q + tk)$-query algorithm finds a verifying set to the verifier with probability at most $\eta$. Then any $Q$-query algorithm computes $t$ disjoint sets $W_1, \dots, W_t$, which are all accepted by the verifier, with probability at most $e^{tk} \cdot \eta^t$.
\end{lemma}

The reader should interpret this as follows. Some algorithm seeks to identify portions of the oracle causing the verifier to accept (which we'll call ``solutions'') by making queries to the oracle. Lemma~\ref{lem:crab-intro} says that if it is hard for query efficient algorithms to find one solution, then it is significantly harder for query efficient algorithms to find many solutions involving pairwise disjoint portions of the oracle; in particular the probability of finding $t$ disjoint solutions will decay exponentially in $t$.

A highly related result is the van den Berg-Kesten-Reimer (BKR) inequality \cite{van1985inequalities, reimer2000proof}, which is concerned instead with the \emph{existence} of many disjoint solutions. In our terminology, the BKR inequality states that if a solution causing the verifier accept exists with probability $\eta$ (over the random choice of oracle), then $t$ disjoint solutions exist with probability at most $\eta^t$. Although originally studied due to its applications to percolation theory, the BKR inequality and its variants have since been applied to several problems in theoretical computer science, including oracle separations in cryptography and relations between complexity measures of Boolean functions \cite{kahn2000dual, smyth2002reimer}.
In contrast to our lemma, the BKR inequality says nothing about solutions that exist but are computationally hard to find (recall that throughout this discussion, computation cost is measured by query complexity).
Our result can be thought of as a kind of ``computational analog'' of the BKR inequality, which the authors believe is of independent interest. For our application of this lemma to vector commitments this algorithmic aspect is crucial; we are interested in events whose probability may be large (or even 1), but with the property that query-efficient algorithms are very unlikely to find a \emph{witness} that the event holds. This situation is ubiquitous in the study of the random oracle model (and query complexity more generally), and hence we are optimistic that it may prove a useful tool for future random oracle algorithms and separations.

Lemma~\ref{lem:crab-intro} also bears a strong resemblance to classical \emph{direct product theorems}, which state that if no $Q$-query algorithm can find a solution to $V^{\calR}$ with probability more than $p$, then for independent random oracles $\calR_1, \dots, \calR_t$, no $Q$-query algorithm can find solutions to $V^{\calR_1}, \dots, V^{\calR_t}$ simultaneously with probability greater than $\eta^t$ \cite{Sha03, drucker2012improved}. To compare this to our setting, we can imagine the $t$ separate oracles $\calR_1, \dots, \calR_t$ as a single random oracle partitioned into $t$ parts; in the direct product theorem we want to bound the probability of finding a complete solution within each part. In Lemma~\ref{lem:crab-intro} we have a more difficult task: we need to bound the the probability of finding $t$ disjoint solutions, where the partition of the oracle into the $t$ parts containing each solution is not fixed in advance, but rather depends on the oracle and the query algorithm.

Unlike the BKR inequality and direct product theorems, which have no dependence on the number of queries made by the verifier, our bound in the disjoint product lemma pays a cost exponential in $tk$. We show that without additional assumptions on the verifier or the parameter regime, this additional cost is necessary:

\begin{lemma}
\label{lem:lb-intro}
For any $\alpha \in (0,1]$ and $t$ sufficiently large, there exist a set $\calX$ and a verifier $V$ that queries a random oracle $\calR: \calX \rightarrow \zo$ in at most $k = \log^{O(1/\alpha)}(|\calX|)$ positions such that
    \[\eta_t \geq \eta_1^t \cdot e^{t k^{1-O(\alpha)}} \]
where $\eta_t$ denotes the maximum probability a $\textsf{poly}(k)$-query algorithm succeeds at finding $t$ disjoint solutions and $\eta_1$ is the maximum probability that a $2^{k^{O(\alpha)}}$-query algorithm succeeds at outputting a single solution.
\end{lemma}

\section{Related work} \label{sec:related-work}

Merkle trees were introduced by Ralph Merkle in his 1979 thesis~\cite[Chapter V.4: Tree Authentication]{Merkle1979}, as a tool for constructing hash-based digital signatures.
Specifically, Merkle describes his tree authentication protocol as solving the problem of committing to a long vector while providing quick authentication of each component.
This is exactly the functionality of a \emph{vector commitment}, though the primitive was formally defined decades later~\cite{catalano2013vector}.
A Merkle tree also predates and provides the weaker functionality of an \emph{accumulator}, also formally defined later~\cite{benaloh1993one}.

\paragraph{Accumulators and vector commitments.}
Accumulators compute a short (typically, $O(\secpar)$ bits) digest of a large set and provide membership witnesses for each element.
A verifier can check these witnesses in time sublinear in the set size.
Accumulators were introduced by Benaloh and de Mare~\cite{benaloh1993one}, who also gave the first algebraic construction, which is RSA-based.
Later constructions offer improved functionality such as non-membership witnesses~\cite{CL02,LLX07,nguyen}.

Vector commitments~\cite{catalano2013vector} extend the functionality of an accumulator to vectors.
A witness certifies that a vector's $i^{\text{th}}$ component is $x_i$, and it is infeasible for an attacker to produce witnesses for two different symbols at the same index.
A long line of work builds vector commitments from a variety of assumptions, seeking to minimize digest size, witness length, and verifier complexity, in addition to improved functionalities such as aggregation of witnesses for multiple indices~\cite{lai2019subvector,tomescu2020aggregatable,gorbunov2020pointproofs,peikert2021vector}.

A further extension of vector commitments that Merkle trees can be adapted to instantiate is the \emph{authenticated dictionary} primitive~\cite{papamanthou2016authenticated,tomescu2019transparency,tomescu2020authenticated,agrawal2020kvac,tyagi2022versa,fiore2023cuckoo}.
This data structure commits to a key-value store, and each witness certifies the value for a key.
Security requires that each key has at most one value.

\paragraph{Applications to SNARGs.}

Merkle trees are a crucial building block of succinct non-interactive arguments (SNARGs) in the random oracle model~\cite{DBLP:conf/tcc/Ben-SassonCS16,DBLP:conf/ccs/AmesHIV17,DBLP:journals/iacr/Ben-SassonBHR18,DBLP:conf/eurocrypt/Ben-SassonCRSVW19}.
Such objects allow a prover to convince a verifier that an instance belongs to a language.
In one popular construction framework, the prover commits to a PCP proof of the statement using a vector commitment, typically a Merkle tree.
The verifier then queries a subset of positions, and the prover returns these PCP symbols along with their witnesses.
An application of Fiat-Shamir makes this non-interactive.

This framework is also applied to interactive oracle proofs (IOPs) rather than PCPs, where similarly the prover commits to oracles using a VC, and the verifier receives VC witnesses for its queries.
Such hash-based SNARGs are among the most practically efficient and are used by Google for privacy-preserving age verification\footnote{The Longfellow system by Frigo and shelat uses Merkle trees: \\\url{https://www.ietf.org/archive/id/draft-irtf-cfrg-longfellow-00.html}, \\\url{https://developers.google.com/wallet/identity/verify/accepting-ids-from-wallet-online}} and by the cryptocurrency Starknet for proving transaction validity.\footnote{Starknet: \url{https://starkware.co/starknet/}}.

Minimizing the verifier complexity is a primary goal of SNARGs.
For constructions in this framework, a rough heuristic for the verifier complexity is
\[
(\text{complexity of VC verification}) \cdot (\text{number of PCP or IOP queries})
\]
Thus, vector commitments with improved verifier complexity are a natural direction toward constructing more efficient SNARGs.
Indeed, Fenzi's talk ``SNARKs in practice'' explicitly poses the open question:\footnote{Fenzi slides: \url{https://gfenzi.io/presentations/practical-snarks.pdf}}
\begin{quote}
Are Merkle trees optimal as VCs in the ROM?
\end{quote}
There, Fenzi also notes that hash-based SNARKs currently secure billions of dollars in blockchains.

Our lower bound says that, within this framework, asymptotically better vector commitments cannot help: Merkle trees are already optimal.
However, our bound does not rule out cleverer combinations of vector commitments and IOPs, or different construction paradigms altogether.
Our overall understanding of the verifier complexity of SNARGs in the ROM is fairly limited.
The strongest existing lower bound~\cite{HNY22} exhibits a large gap with constructions and further assumes rETH and a nonstandard security notion called salted soundness.
Proving stronger ROM-SNARG lower bounds under standard soundness is an interesting open question.

\paragraph{Lower bounds for authenticated data structures.}
There are many lower bounds in the literature, which differ from ours in terms of the data structure (e.g., accumulator), the model (e.g., the random oracle model), or the quantity lower bounded.
The most closely related bound to ours is by Tamassia and Triandopolous~\cite{DBLP:conf/icalp/TamassiaT05}, who show that a restricted class of ``DAG hash-based'' authenticated data structures require verifier complexity $\Omega(\log n)$.
This class contains Merkle trees but is fairly narrow, consisting of constructions whose oracle usage forms a DAG; verifier complexity is measured as a function of the number of queries and their in-degree.\footnote{This complexity measure differs from ours, which simply counts the number of verifier queries, and is why their bound does not contradict high-arity Merkle trees making $\log_{\theta} n$ queries.}

Others bound the number of times witnesses must change as elements are added to or deleted from an accumulator~\cite{mahmoody2022lower,christ2022limits,DBLP:conf/crypto/BonneauCCK25,10.62056/av7t7ta5v} but say nothing about the complexity of verifying these witnesses.
There are similar lower bounds on the complexity of maintaining memory checkers as the memory changes~\cite{dwork2009efficient,boyle2024memory,wang2023locality}.
Like our result, many of these lower bounds are shown using compression arguments, though they bound quantities orthogonal to verifier complexity.

\paragraph{Lower bounds in the random oracle model.} %
Lower bounds in the random oracle model have been studied for several cryptographic primitives. Impagliazzo and Rudich \cite{IR89} and  Barak and Mahmoody \cite{BM17} established a limitations on key agreement from random oracles. Barak and Mahmoody \cite{BM07}   proved a query-complexity lower bounds for one-time signatures. More recently, \cite{CHMY26} improved the lower bound for signatures schemes with non-adaptive verifier, assuming a stronger security notion called \emph{salted soundness}.
This property was introduced for the sake of these lower bounds and, roughly speaking, requires security even against attackers who can resample the random oracle.
Chiesa and Yogev \cite{CY20} proved barriers to succinct arguments with low verifier query complexity, and Haitner, Nukrai, and Yogev \cite{HNY22} proved lower bounds on the proof length of SNARGs and subvector commitments under rETH. %

Subvector commitments appear to be far stronger objects than vector commitments, enabling efficient proofs of any \emph{subset} of indices. In contrast, our work establishes an unconditional lower bound on the verifier query complexity of vector commitments.
Roughly translating their result to vector commitments, they show that each witness must be at least $\secpar/\log n$ bits long, assuming exponential-query security, salted soundness, and rETH.
This bound is smaller than the length of a \emph{single} random oracle input in our setting.

\paragraph{Amplification theorems.} Hardness amplification via direct-product theorems has been
extensively studied in theoretical computer science
\cite{Raz98,Hol09,IJKW10,Sha03,drucker2012improved,braverman2013direct}
and cryptography
\cite{canetti2005hardness,dodis2009security,haitner2009parallel,chung2010parallel}.
In particular, Canetti,  Halevi and  Steiner \cite{canetti2005hardness}
established hardness amplification for weakly verifiable
puzzles through independent parallel repetitions.
Note that in our disjoint-product lemma the instances are not independent.

\section{Technical Overview} \label{sec:techo}

The heart of the proof is a \emph{compression argument}~\cite{GT00}.
This argument involves two parties: \emph{Alice}, who wishes to transmit a bit-vector to \emph{Bob} using as few bits as possible.
To obtain our lower bound, we will present a communication scheme where Alice uses a vector commitment to succinctly encode her vector.
We will show that if this vector commitment's opening proofs involve too few queries, Alice can transmit her vector in \emph{fewer bits than should be information theoretically possible}.

Towards our communication scheme, let us start by examining how a vector commitment (VC) could be useful.
Suppose we have a VC with $\secpar$-bit commitments\footnote{We make this assumption for the sake of presentation in the technical overview, but our ultimate bound allows for arbitrary commitment length.} to $n$-bit vectors, for $n$ much bigger than $\secpar$.
This gives Alice a way to compress vectors into shorter messages: for a vector $x \in \zo^n$, its commitment $c \gets \Com(x)$ is only $\secpar$ bits long.
So let Alice begin by sending $c$ to Bob.

Now, Bob must somehow interpret this commitment $c$. The VC naturally suggests a way to ``decode'' a commitment, using the verifier.
In particular, for each position $i \in [n]$:
\begin{itemize}
    \item \textbf{Completeness} ensures that using $x$, one can generate a witness $\pi_i$ that causes the verifier to accept the correct bit: $\Ver(c, i, x_i, \pi_i) = 1$.
    We assume (without loss of generality for deterministic verifiers)\footnote{Shown in (\Cref{fact:verifier-transform}).} that the witness consists of at most $k$ strings $(q_1, a_1), \ldots, (q_k, a_k) \in \zo^{\theta + \secpar}$, which are alleged random oracle input-output pairs.
    \item \textbf{Position binding} requires that it is infeasible for any polynomial-time algorithm to find $\pi'_i$ causing the verifier to accept the wrong bit: for all probabilistic polynomial query $\adv$, generating $\pi'_i$, $\Ver(c, \bar{x_i}, \pi'_i) \neq 1$ with overwhelming probability.
\end{itemize}
Therefore, to decode, Bob just needs to find an accepting witness $\pi_i$ for every index $i \in [n]$.
As long as Bob makes a polynomial number of queries, position binding ensures that he will never stumble across an incorrect witness, and the bits accepted by the verifier are exactly those Alice chose.

The crux of the scheme is how Bob actually obtains these witnesses.
It is not clear how he can find a witness knowing the commitment alone; in fact, for a standard Merkle tree, doing so requires exponentially many queries to the random oracle.
Of course, our communication scheme assumes a VC with shorter witnesses than Merkle trees, but it is not obvious how this shortness helps Bob.
Alice therefore must help Bob by sending some additional information, but she cannot afford to send all $n$ of these $(k \theta)$-bit witnesses.

Alice's task is much easier if the $n$ witnesses share many queries.
This is true for standard Merkle trees: just the $2n$ queries in the tree comprise all $n$ of the $(\log n)$-query witnesses.
But Alice is given an \emph{arbitrary} VC, and its witnesses must have \emph{even higher} overlap than Merkle trees: Alice can send Bob only $O(n/\theta)$ queries in fewer than $n$ bits.

Unfortunately, it is easy to come up with a contrived VC whose witnesses have many disjoint queries.
Just append to each witness several randomly chosen dummy queries that the verifier ignores.
But perhaps this seems like a silly counterexample, as Bob can fill in these queries on his own (with any value, since they are ignored), without Alice needing to send anything.
Therefore, we can still hope that, setting aside such trivial queries that Bob can find on his own, \emph{there are at most $O(n/\theta)$ nontrivial queries among the $n$ witnesses}.

Amazingly, this hope is true for \emph{all} VCs where the verifier makes $o(\log_\theta n)$ queries to the random oracle.
Specifically, suppose Bob maintains a table of queries he has made and their corresponding answers.
Let $p$ be some hardness parameter that is polynomially bounded in $n$, and say at some point during the protocol that the table is $p$-hard if Bob's best $p$-query algorithm has at most $1/p$ probability of finding a verifying witness for any component he does not yet know and $p$-easy otherwise. Here, we fix the commitment and Bob's current oracle table, and sample the remaining oracle answers uniformly at random.

Note that whenever Bob's table is $p$-easy, there exists a set of $p$ queries to the oracle that he can make to reveal a new bit $x_i$ with probability at least $1/p$, allowing him to make progress by himself. As such, the case we are worried about is where, after Alice has sent Bob some queries, the table is $p$-hard and many honest witnesses have disjoint subsets of queries missing from the table. We call each uncertified index's missing honest queries a \emph{residual set}. Then we are stuck: Bob cannot fill in the witnesses himself, and disjointness means that sending all of these residual sets would cost Alice too many bits.

Fortunately, we show that there are few disjoint residual query sets as a function of the number of bits Alice has sent thus far.
Specifically, the probability that Alice's $\ell$-bit message leaves Bob with a $p$-hard table and $\ell + t$ disjoint residual sets to fill in is at most $2^{-t}$.
Intuitively, if many disjoint sets are individually hard for Bob to find, \emph{finding them together is too hard even for the honest committer}.

At first glance, it is difficult to relate Bob's hardness in finding the witnesses to Alice's.
Considering Merkle trees again, there, Bob must solve a hard hash-preimage problem to find a witness given a commitment.
For Alice, finding witnesses is easy, as she constructs the Merkle tree herself.
The asymmetry is exactly that Alice chooses the commitment and her message as a function of the random oracle responses she receives; in contrast, Bob is given a commitment and then must find the witnesses.

That said, we are able to prove this by giving a hardness amplification lemma for finding disjoint solutions to a random oracle problem. Akin to classical direct product theorems, which related solving a problem over one oracle to solving it simultaneously over many independent oracles, we prove that it is exponentially harder to find $t$ disjoint solutions than a single solution. Intuitively, this should be expected as disjoint solutions ought to be independent of one another.

With this lemma in hand, given an $\ell$-bit message $\alpha$ consisting of a commitment and queries fixed independently of the random oracle, the best $O(p)$-query strategy finds $t$ $p$-hard disjoint witnesses with probability at most $2^{\Omega(-t)}$.
We then relate this to the honest committer's execution using a union bound.
That is, for an advice string $\alpha$ let $\Bad(\alpha)$ denote the event that Bob's table $T$ is $p$-hard and the honest execution produces witnesses with disjoint residual sets under $T$ in $O(p)$ queries.
Our disjoint product lemma says exactly that $\Pr[\Bad(\alpha)] \leq 2^{-t}$.
Now, the probability that the honest committer's execution produces these disjoint witnesses is
\begin{align*}
    \sum_{\alpha \in \{0,1\}^\ell}\Pr[\text{Alice's message is } \alpha \land \Bad(\alpha)] \leq 2^{\ell} \cdot \Pr[\Bad(\alpha)] \leq 2^{\ell - t}.
\end{align*}

We believe this \emph{disjoint product lemma} (\Cref{lem:crab}) is of independent interest, and we give a technical overview of it and its proof in \Cref{sec:to-crabs}.

We now present the communication protocol in greater detail. Although Alice is the only party sending messages, and therefore she could send her message all at once, it will be helpful to view the protocol as occurring in rounds.
Places where Alice sends messages are indicated in \text{bold}.

\begin{description}
    \item[Setup phase.] Alice and Bob agree on a hardness parameter $p$ that is slightly more than the number of queries made by the honest committer.

    \item[Vector sampling.] Alice samples her vector, a uniformly random vector $x \in \{0,1\}^n$, whose length $n$ is a large polynomial in the security parameter $\secpar$. Her goal is to transmit $x$ to Bob, by sending fewer than $n$ bits from this point on.

    \item[Sender round 0.] Alice generates her commitment and witnesses: $c, \pi_1, \ldots, \pi_n \gets \Com(x)$. Each witness $\pi_i$ consists of $k$ queries $q_1, \ldots, q_k \in \zo^\theta$.
    \textbf{Alice sends $c$ to Bob}.

    \item[Receiver round 0.] Bob initializes an empty table $T$, where he will record the queries and responses he receives from the random oracle.

    Bob then checks if $T$ is $p$-easy given all information he has seen thus far: there is a $p$-query strategy that succeeds in finding a witness for some uncertified index with probability at least $1/p$.\footnote{Bob can compute the success probability (over the randomness of the random oracle's responses) of any strategy without querying the random oracle. This might take exponential time, but recall that all complexity is measured only in the number of random oracle queries.}
    If so, Bob runs this good strategy and records all queries' input-output pairs in his table.
    If any such queries comprise a verifying witness, i.e.,
    \[
    \Ver^{\calR}(c, i, b, (q_{j_1}, \ldots, q_{j_k})) = 1
    \]

    for some inputs $(q_{j_1}, \ldots, q_{j_k})$ in Bob's table, we say $i$ is \emph{certified}.
    He then checks whether this updated table is $p$-easy given the additional queries he has made and repeats.

    A simple bound ensures that with all but negligible probability, in $\Theta(p^2 \cdot n^2)$ queries Bob either reconstructs the entire vector or arrives at a $p$-hard table. We call this procedure ``hardening Bob's table.''

    \item[Sender round 1.] Alice has been following along, making the same oracle queries as Bob, who is deterministic from her perspective. Therefore, she knows $T$ and which indices Bob has certified.

    Alice now examines the uncertified indices $S \subseteq [n]$ and greedily selects a maximal set of honest witnesses whose remaining oracle queries are pairwise disjoint; call their indices $S^* \subseteq S$.
    Maximality ensures that every \emph{other} uncertified honest witness shares at least one query with this maximal set.

    Therefore, by \textbf{sending all remaining queries of the honest witnesses of $S^*$}, Alice certifies $S^*$ \emph{and ensures that all uncertified indices have at least one query from their honest witness in $T$}.
    In general, we will preserve this invariant: after round $i$, every index will either be certified or have at least $i$ queries from its honest witness in $T$.

    Sending these queries costs Alice up to $k \cdot \abs{S^*} \cdot \theta$ bits.
    Therefore, \textbf{Alice aborts if the number of indices whose sets of remaining queries are pairwise disjoint is too large}: $\abs{S^*} \geq \ell + \secpar$, where $\ell$ is the number of bits Alice has sent before the current round.
    Our disjoint product lemma ensures that Alice aborts with probability at most $2^{-\secpar}$.

    \item[{Receiver round $i \in [k]$}.] Bob receives some number of inputs in $\zo^\theta$ from Alice, queries them to the random oracle, and adds them and their outputs to his table.
    He notes which indices they certify; \textbf{if Bob certifies the entire vector he outputs the bits the verifier accepts}.
    Otherwise, Bob hardens his table.

    \item[Sender round $i+1$.] Alice examines the uncertified indices $S \subseteq [n]$ and greedily selects a maximal set of honest witnesses whose remaining oracle queries are pairwise disjoint; call their indices $S^* \subseteq S$. \textbf{Alice aborts if $\abs{S^*} \geq \ell + \secpar$}, where $\ell$ is the number of bits Alice has sent before the current round.

    Otherwise, Alice \textbf{sends all remaining queries of the honest witnesses of $S^*$}, Alice certifies $S^*$ \emph{and ensures that all uncertified indices have at least one query from their honest witness in $T$}.
\end{description}

\paragraph{Bob succeeds in reconstruction.} Recall our invariant: after round $i$, every index is either certified or has at least $i$ queries from its honest witness in $T$.
Therefore, after $k$ rounds, all indices will be certified (unless its honest witness does not verify, which constitutes a completeness violation).
Except with this negligible probability of completeness error, the only way Bob can fail is if he outputs the incorrect bit at some position.
Since Bob only outputs certified bits, this would exactly mean he has found a verifying witness for an incorrect bit.
Since Alice knows the honest witnesses, Alice and Bob together produce witnesses for $x_i = 0$ and $x_i = 1$.

The last thing we must check before invoking binding is that Alice and Bob together make polynomially many queries.
Alice only runs the honest commitment algorithm, then makes the same queries as Bob.
In each of $k$ rounds, Bob makes $O(p^2 \cdot n^2)$ queries to harden his table, plus $\abs{S^*} \leq n + \secpar$ queries from Alice's set.
This is overall polynomial, and position binding holds.

\paragraph{Counting the number of bits Alice sends.}
By the invariant, the protocol terminates at the end of round $k$.
In each round $i = 0, \ldots, k$, let $\ell_i$ denote the number of bits Alice sends in that round.
Since Alice only sends the commitment in round 0, $\ell_0 = \secpar$.
Assuming Alice does not abort, for subsequent rounds $i \geq 1$ we have the following relation:
\[
\ell_i \leq k \cdot \theta \cdot \left(\secpar + \sum_{j = 0}^{i} \ell_j \right).
\]
Writing $L_i = \secpar + \sum_{j = 0}^{i} \ell_j$, we have $L_0 = 2\secpar$ and $L_i \leq (1 + k \theta) L_{i-1}$.
Therefore, Alice's total communication is at most
\[
2 \secpar(1 + k \theta)^k \leq 2e\secpar(k\theta)^k.
\]

\paragraph{Arriving at a contradiction.} Since Bob successfully decodes Alice's random $n$-bit vector, Alice must have sent at least $n - 1$ bits.\footnote{This is $n-1$ rather than $n$ bits because we must account for the negligible probability of incorrect reconstruction due to completeness or binding errors.}
Therefore, we obtain the bound:
\[
2e\secpar(k\theta)^k \geq n-1, \text{ simplifying to } k = \Omega(\log_{\theta} n).
\]

\begin{figure}[t]
  \centering
  \includegraphics[width=0.6\textwidth]{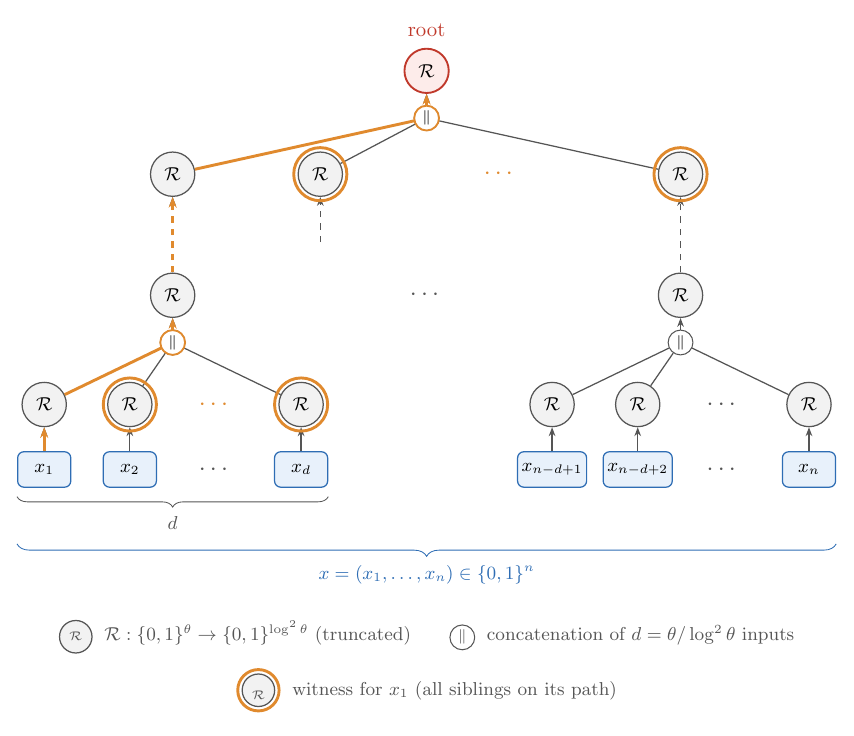}
  \caption{A Merkle tree of arity $d=\theta/\log^2\theta$. Oracle outputs
    are truncated to $\log^2\theta$ bits, so $d$ children fill one
    $\theta$-bit query. Highlighted in orange is the witness of $x_1$,
    consisting of the siblings along its path to the root.}
  \label{fig:merkle-dary}
\end{figure}

\subsection{A Tight Upper Bound} \label{sec:high-arity}
We remark that our lower bound is matched by high-arity Merkle trees (shown in \Cref{fig:merkle-dary}) in the natural regime, where $\theta \geq \secpar$ (i.e., the hash function is compressing).
Truncate all random oracle outputs to $\ell = \log^2 \theta$ bits, and let the arity be $t = \theta/\ell$.
Because of our output truncation, we can pack $t$ siblings' outputs into a single $\theta$-bit query to the random oracle.

For a vector of length $n$, the depth of this tree is \[
\log_t n = \frac{\log n}{\log \theta - 2 \log \log \theta} = \Theta\left( \frac{\log n}{\log \theta}\right).
\]

Therefore, each opening requires $\Theta(\log_{\theta} n)$ verifier queries, matching our lower bound.

\subsection{The Disjoint Product Lemma} \label{sec:to-crabs}

It now remains to discuss the proof of the disjoint product lemma, which recall roughly seeks to argue that if no $Q$-query algorithm can find a verifying set with probability more than $\eta$, then any $Q$-query algorithm finds $t$ disjoint verifying sets with probability at most $e^{tk} \eta^{t}$.

Before discussing the proof, we first discuss why one might suspect such a statement to be true. Roughly, this comes from the fact that under a product distribution, we expect the disjoint solutions to essentially behave as though they come from independent oracles. In such a direct product setting, knowing a solution to one of the oracles should not help find solutions to another oracle. It turns out that, while it yields worse bounds, we can in fact formally reduce from the disjoint product lemma to direct product results as we'll see in the upcoming discussion.

Further intuition comes from the van den Berg-Kesten-Reimer (BKR) inequality \cite{van1985inequalities, reimer2000proof}, which roughly states that for a product distribution over $\zo^n$ and DNFs $f$ and $g$, the probability that a point $x$ drawn from the distribution satisfies a term in $f$ and a term in $g$ that are disjoint is at most the product of the probability that $f$ is satisfied by $x$ and the probability $g$ is satisfied by $x$. Intuitively, $f$ and $g$ are not any more likely to be satisfied for disjoint reasons than what independence would suggest. We can then think of our result as an algorithmic cousin of this inequality, relating the probability of finding disjoint witnesses rather than the probability that they exist.

\paragraph{A direct product lemma.} As a starting point, we begin by describing how to prove a simpler direct product lemma. In this setting, we wish to show that if no $Q$-query algorithm can find a verifying set under $V^{\calR}$ with probability more than $\eta$, then any $Q$-query algorithm with access to independent random oracles $\calR_1, \calR_2, \dots, \calR_t$ can compute verifying sets for all of $V^{\calR_1}, \dots, V^{\calR_t}$ with probability at most $\eta^t$. To do this, we show how to take an algorithm $\alg$ that outputs solutions to $V^{\calR_1}, \dots, V^{\calR_t}$ with probability $\delta$ and compile it into an algorithm that outputs a single verifying set for $V^{\calR}$ with probability at least $\delta^{1/t}$.

Indeed, we can naturally get such an algorithm by taking the true random oracle $\calR$, choosing a random $r \in [t]$, and setting $\calR_r = \calR$. The naive argument would then be to simulate $\alg$ by flipping our own independent coins to generate values from the remaining oracles $\calR_i$ for $i \in [t] \setminus \{r\}$. That said, such an approach may not work as $\alg$'s success on $\calR_r$ may occur if and only if it finds a verifying set for the remaining oracles $\calR_i$. For instance, $\alg$ could choose to output nonsense when it sees it won't succeed on all the oracles, even if it found solutions to some of the oracles. To circumvent this issue, we use the fact that we can choose the values of the oracles $\calR_i$ for $i \not = r$ and do not need to set them randomly. Indeed, we should always answer our queries with respect to $\calR_i$ for $i \not = r$ in such a way as to direct ourselves towards outcomes where we output $t$ solutions. In particular, we can choose the random oracle values of $\calR_i$ for $i \not = r$ to be uniform conditioned on the event that we output a solution to all $t$ oracles.\footnote{Arguably, it is more natural to choose the values of the oracles so as to maximize the probability over $\calR$ of outputting solutions to each of $V^{\calR_1}, \dots, V^{\calR_t}$. That said, in future settings, it is more informative to answer queries in a manner that is more invariant of $r$.} Implementing such a procedure, we can then prove that this scheme outputs a verifying set to $V^{\calR}$ with probability at least $\delta^{1/t}$.

\textbf{Reducing from disjoint products to direct products.} We now turn to understanding how to handle the disjoint product lemma, wherein there is only a single oracle. Our approach will again be to take an algorithm $\alg$ that outputs $t$ disjoint verifying sets, $W_1, W_2, \dots, W_t$, with probability $\delta$ and show how to compile this into a new algorithm that outputs a single verifying set with probability at least $e^{-k} \delta^{1/t}$.

Notably, at some quantitative loss, we can in fact reduce from the previous direct product setting. In particular, let $\calR_1, \dots, \calR_t$ denote $t$ random oracles. Randomly color each element in $\calX$ with a color from $[t]$ and create an oracle $\calR$ by setting $\calR(q) = \calR_{c(q)}(q)$ where $c(q)$ is the random color of $q$. We can then run $\alg$ on $\calR$. Since $\calR$ is distributed according to a standard random oracle, we output $t$ disjoint verifying sets $W_1, \dots, W_t$ with probability $\delta$. Moreover, since the coloring was random and independent of $\calR_1, \dots, \calR_t$, we have that the probability that for all $i \in [t]$ all the elements in $W_i$ are labeled with $i$ is at least $t^{-tk}$. Thus, we solve the direct product problem with probability at least $t^{-tk} \delta$. We can then use our previous scheme to output a solution with probability at least $t^{-k} \delta^{1/t}$, which falls slightly short of our desired target success probability.

\paragraph{Improving the simulation.} We now turn to describe how to get the final disjoint product lemma. It's not hard to see that the above argument is lossy because of the coloring argument, as when $t \gg k$ we don't expect any of the sets $W_1, \dots, W_t$ to be monochromatic. Thus to get the sharper simulation bound, we need to remove the coloring scheme from the above compiled algorithm.

Roughly, the current simulation samples a value $r \sim [t]$ and sets $\calR_r = \calR$ and then tries to maximize the probability that we find a solution in every one of the color classes. Given that the colors are not inherent to the problem and are causing a quantitative loss, it's natural to remove them and simply answer all our fictitious oracle queries in such a way as to end at a uniformly random outcome with $t$ disjoint solutions $W_1, \dots, W_t$. Since all the oracles $i \not = r$ may as well answer queries in the same way, we'll simply assume there is only one fictitious oracle, which we call $\ideal$.

With this modification, our compiler now simulates $\alg$ and at each step samples a random variable $i \sim [t]$, which corresponds to the color of each input point. If it is equal to $r$, then we answer the query with the true random oracle. Otherwise, we answer with $\ideal$. From this perspective the loss in the previous bound can be viewed as coming from $\calR$ and $\ideal$ being completely unrelated. To fix this, we sample $\ideal$ conditioned on $\calR$, using a coupling that preserves the marginal distributions of both oracles while maximizing their probability of agreement on each query.

This then leads us to our final simulation: We answer all of our queries according to $\ideal$, which will agree non-trivially with $\calR$ by the coupling. By design, we end with $t$ disjoint sets $W_1, \dots, W_t$ that are accepted by the verifier under $\ideal$. While it takes some work to compute, we can then show that a uniformly random set $W_i$ satisfies the verifier $V^{\calR}$, by virtue of $\calR$ and $\ideal$ agreeing on the queries in $W_i$, with probability at least $e^{-k} \delta^{1/t}$. If we assume that no $Q$-query algorithm could output a solution to the verifier with probability better than $\eta$, it then follows that the probability of outputting $t$ disjoint solutions satisfies $\delta \leq e^{tk} \eta^t$, as desired.

\paragraph{Lower bounds.} For the lower bound, our key construction, due to ChatGPT Astra, will be parameterized by $t$. It shows that there is a verifier such that we can find $t$ disjoint solutions with probability at least $1/2$ using $\textsf{poly}(t)$ queries, but no algorithm making $2^{O(t)}$ queries can find a single solution with probability better than $1/2 + o_t(1)$. Given such a scheme, the lower bound then follows by applying a direct product lemma. In particular, we consider $\ell$ oracles $\calR_1, \dots, \calR_\ell$ and have our verifier $V^{\otimes \ell}$ accept if and only if $V^{\calR_i}$ accepts for all $i$. Roughly, this then yields an algorithm making $\textsf{poly}(t \ell)$ queries that outputs $t$ disjoint solutions with probability at least $2^{-\ell}$, while no algorithm making $2^{O(t)}$ queries can output a single witness with probability greater than $(1/2 + o_t(1))^\ell$.

We now turn to describing how to construct the instance where it is roughly just as hard to produce a single witness as many disjoint ones. Recall that in our setting, we assume the random oracle $\calR$ simply outputs a single random bit for each point in the domain. Our approach aims to correlate our success with an event. In particular, we build a ``core'' of $N = t^8$ elements $z_1, \dots, z_N$. We would like to then have it be easy to output many disjoint solutions when $\calR(z) := \calR(z_1), \dots, \calR(z_N)$ has at least $\frac{N}{2}$ zeros and hard to find even a single solution when $\calR(z)$ has at least $N/2 + 4t^3$ ones. Towards this, the verifier will take as input a string $x \in \zo^N$ that it expects to satisfy $x = \calR(z)$. That said, the verifier will reject any string $x$ with more than $N/2$ ones.

Now, we cannot check that $x_j = \calR(z_j)$ for all indices as our verifier needs to have $t$ witnesses with disjoint sets of queries. As such, we partition $[N]$ into $t$ pieces, $P_1, P_2, \dots, P_t$. The verifier then receives an index $i$ and only checks that $x_j = \calR(z_j)$ for $j \in P_i$. While this is now amenable to disjoint witnesses, using a fixed partition will make finding a single witness too easy as an algorithm can simply set $x_j = \calR(z_j)$ for $j \in P_i$ and $x_j = 0$ elsewhere to get an accepting witness.

To circumvent this, we will associate a sequence of points in $\calX$, $(x,y_1), (x,y_2), \dots, (x,y_{t^2})$, to each string $x \in \zo^N$. This then associates a random string to each $x$ using the random oracle, $r(x) = \calR((x,y_1)) \calR((x,y_2)) \dots \calR((x,y_{t^2})) \in \zo^{t^2}$, which in turn will essentially be used as a seed to generate a random partition $P_1(r(x)), \dots, P_t(r(x))$. This will then prevent the algorithm from easily finding a low Hamming weight $x$ that agrees with $\calR(z)$ when $\calR(z)$ has many ones.

Unfortunately, we have now seemingly entered into a chicken-and-egg scenario as the verifier cannot always query the oracle on $(x,y_1), (x,y_2), \dots, (x,y_{t^2})$ to figure out the random seed $r(x)$ if we want disjoint witnesses. That said, by a concentration argument, there are very few random seeds that would result in a bad partition where $x$ and $\calR(z)$ agree on some equivalence class but disagree on at least $4t^3$ indices. Given this, the verifier can get by only checking that $s$ and $r(x)$ agree on a small number of indices. In particular, our final verifier takes as input $(x,s,i)$ with $x \in \zo^N$, $s \in \zo^{t^2}$, and $i \in [t]$. First and foremost, it rejects if $x$ has more than $N/2$ ones. Next, it rejects if $x_j \not = \calR(z_j)$ for some $j \in P_i(s)$, where $P_i(s)$ is the $i$th piece of the random partition associated to the string $s$. Finally, we must check that we used the correct partition, which is accomplished by simply checking that $s_j = \calR((x,y_j))$ for all indices $j \in (t(i-1), ti]$.

It's now easy to check that when $\calR(z)$ has at least $N/2$ zeros, we can simply set $x = \calR(z)$, $s = r(x)$ and take $(x,s,i)$ for $i \in [t]$ to be our disjoint witnesses. On the other hand, due to the random partitions, no algorithm can output a single solution with probability better than $o_t(1)$ when $\calR(z)$ has at least $N/2 + 4t^3$ ones unless it makes $2^{\Omega(t)}$ queries. As such, we only expect to output a solution when there are fewer than $N/2 + 4t^3$ ones, which happens with probability roughly $1/2 + O(t^3/\sqrt{N}) = 1/2 + O(1/t)$.\footnote{We remark that it's not obvious to see directly, but there indeed are solutions when $\calR(z)$ has high Hamming weight. In particular, applying the BKR inequality implies that there must exist a witness with probability at least $1 - O(1/t)$; it simply requires many queries to find.}

\section{Preliminaries}
\paragraph{Notation.} For an integer $n$, let $[n]:=\{1,\ldots,n\}$.
We denote by $\{0,1\}^n$ the set of $n$-bit strings and
by $\{0,1\}^*$ the set of all finite binary strings.
For a string $x$, let $|x|$ denote its bit length, and
for a set $S$, let $|S|$ denote its cardinality.
We write $x\leftarrow D$ to denote sampling $x$
according to a distribution $D$, and
$x\gets S$ for uniform sampling from a finite set $S$.

Throughout, $\lambda$ denotes the security parameter.
A function $\nu:\mathbb{N}\to[0,1]$ is negligible (denoted by $\nu=\negl$) if
for every polynomial $p$, there exists $\lambda_0$
such that $\nu(\lambda)<1/p(\lambda)$ for all
$\lambda\ge\lambda_0$.
An event occurs with overwhelming probability if
its probability is at least $1-\negl$.

We write $\log$ to denote the logarithm base 2 and $\ln$ to denote the logarithm base $e$.

\paragraph{Random oracle model.} We work in the random oracle model, where $\calR: \zo^\theta \to \zo^\secpar$ is a truly random function that can be queried via black-box access. We assume that $\theta$ and $\secpar$ are polynomially related. All complexity is measured in the number of queries made to the random oracle.

A query-answer table $T$ is a set of pairs $(q,a)$.
We write $\dom(T)$ for the set of queries appearing
in $T$, and say that an oracle $R$ is consistent with
$T$ if $R(q)=a$ for every $(q,a)\in T$.
\subsection{Vector commitments}
A vector commitment is a tuple of algorithms $\Sigma = (\Com, \Ver)$:
\begin{itemize}
    \item $\Com^{\calR}(x) \to c, \pi_1, \ldots, \pi_n$: takes as input a vector $x \in \{0,1\}^n$. It outputs a commitment $c$ and witnesses $\pi_1, \ldots, \pi_n$.
    \item $\Ver^{\calR}(c, i, x_i, \pi_i) \to \{\true, \false\}$: is a \emph{deterministic} oracle algorithm that takes as input a commitment $c$, an index $i$, a value $x_i \in \{0,1\}$, and a proof $\pi_i$.
    It outputs $\true$ if and only if the $i^{\text{th}}$ component of the vector underlying $c$ is indeed $x_i$.
\end{itemize}

Efficiency requires that the number of queries made by $\Com$ and $\Ver$ must be upper bounded by a polynomial in $\secpar$ and $n$.
A vector commitment must satisfy completeness and position binding:

\begin{definition}[Completeness]
    A vector commitment satisfies completeness if there
    exists a negligible function $\negl$
    such that for every polynomially bounded
    $n=n(\secpar,\theta)$ and every $x\in\{0,1\}^n$,
    \[
    \Pr_{\calR,\Com}\left[
        \exists i\in[n]:
        \Ver^{\calR}(c,i,x_i,\pi_i)=\false
        \;:\;
        (c,\pi_1,\ldots,\pi_n)\gets\Com^{\calR}(x)
    \right]\leq \negl.
    \]
\end{definition}

\begin{definition}[Position binding]
    A vector commitment satisfies position binding if,
    for every computationally unbounded adversary
    $\adv$ making polynomially many random oracle queries
    and every polynomially bounded vector length
    $n=n(\theta,\secpar)$,
    \[
    \Pr_{\calR,\adv}\left[
    \begin{array}{l|r}
    \Ver^{\calR}(c,i,b,\pi)=\true
    \ \land
    &
    (c,i,b,b',\pi,\pi')
    \gets \adv^{\calR}(1^\secpar,1^n)
    \\[2pt]
    \Ver^{\calR}(c,i,b',\pi')=\true
    &
    b\neq b'
    \end{array}
    \right]
    \leq \negl.
    \]
    That is, no polynomial-query adversary can produce
    a commitment and accepting openings for two different
    bits at the same position, except with negligible
    probability.
\end{definition}

\begin{fact} \label{fact:verifier-transform}
    Any vector commitment with a deterministic verifier making at most $k$ queries per witness can be converted to one where each witness consists of only these queries, and the verifier makes exactly these queries nonadaptively.
    That is, for every $i \in [n]$, the new witness takes the form
    \[
    \pi_i' = ((q_1, a_1), \ldots, (q_k, a_k)) \in \zo^{(\theta + \secpar)k}
    \]
    and the new verifier ${\Ver'}^{\calR}(c, i, x_i, \pi_i')$ queries all $q_j$'s in a single round.
    The new committer makes at most an additional $kn$ queries.
\end{fact}

\begin{proof}
    Suppose we have a vector commitment $\Sigma = (\Com, \Ver)$ with arbitrary witnesses $\pi_i$ of length $\ell$ and a deterministic verifier $\Ver$ that makes at most $k$ random oracle queries $q_1, \ldots, q_k$ given $(c, i, x_i, \pi_i)$.

    We define $\Sigma' = (\Com', \Ver')$ as follows:
    \begin{itemize}
        \item On input $x$, $\Com'$ computes $(c, \pi_1, \ldots, \pi_n) \gets \Com(x)$. For each $i \in [n]$, let $q^{(i)}_1, \ldots, q^{(i)}_k$ be the queries made by $\Ver(c, i, x_i, \pi_i)$.
        Let
        \[
            \pi'_i = ((q^{(i)}_1, \calR(q^{(i)}_1)), \ldots, (q^{(i)}_k, \calR(q^{(i)}_k)))
        \]
        The new committer $\Com'$ sets $c' = c$ and outputs $(c', \pi'_1, \ldots, \pi'_n)$.
        \item On input $(c', i, b, \pi'_i)$, the new verifier $\Ver'$ outputs 1 if and only if
        \begin{multline*}
        \calR(q^{(i)}_j) = a^{(i)}_j \text{ for all } j \in [k] \text{ and } \exists \pi_i \in \{0,1\}^\ell \text{ such that } \Ver^{\pi_i'}(c', i, b, \pi_i) \text{ queries only } \pi_i' \text{ and outputs }1.
        \end{multline*}
        Here, the notation indicates that the oracle queries of $\Ver$ are answered using $\pi_i'$.
        By construction, this new verifier still makes $k$ queries.
    \end{itemize}

    \paragraph{Completeness.} We'll show if $\Ver$ accepts $\pi_i$, $\Ver'$ accepts $\pi'_i$.
    Suppose that $\Ver$ accepts $\pi_i$; then $\pi_i$ itself causes $\Ver(c', i, x_i, \pi_i)$ to query exactly $q^{(i)}_1, \ldots, q^{(i)}_k$. $\Ver$ then accepts when the oracle responds with the $a^{(i)}$'s.

    \paragraph{Binding.} We'll now show that a binding violation for $\Sigma'$ can be converted into a binding violation for $\Sigma$ with no additional queries. Suppose that $\Ver'$ accepts $\pi'_i$ for $b \neq x_i$.
    Then the random oracle $\calR$ matches $\pi'_i$, and there exists $\pi_i$ such that $\Ver(c, i, b, \pi_i)$ queries only the inputs of $\pi'_i$ and accepts whenever the outputs match $\pi'_i$.
    Therefore, $\pi_i$ causes $\Ver$ to accept the incorrect bit.
    Given $\pi'_i$, one can find $\pi_i$ by searching over all $\ell$-bit strings, making no queries to the random oracle.
\end{proof}

\subsection{Cryptographic accumulators}

\begin{definition}[Accumulator \cite{benaloh1993one}\label{def:accumulator}]
An \emph{accumulator} for a data universe $\calU$ is a tuple of algorithms $\Sigma = (\Acc, \Ver)$:
\begin{description}
    \item[$\Acc^{\calR}(S) \to c, (\pi_1, \ldots, \pi_{\abs{S}})$:] $\Acc$ is a (possibly randomized) polynomial-time algorithm that takes as input a subset $S\subseteq U$.
    It returns a \emph{commitment} $c$ and a \emph{membership witness} $\pi_i$ for each $s_i \in S$.
    \item[$\Ver^{\calR}(c, u, \pi) \to \{\true, \false\}$:] $\Ver$ is a deterministic algorithm that takes as input a commitment $c$, an element $u \in \calU$, and a membership proof $\pi$.
    $\Ver$ should output true if and only if $\pi$ is a valid proof of membership of $u$ in the set that $c$ accumulates.
\end{description}
\end{definition}

Efficiency requires that the number of queries made by $\Com$ and $\Ver$ must be upper bounded by a polynomial in $\secpar$ and $n = \abs{S}$.
An accumulator must satisfy \emph{soundness} and \emph{correctness}.

\begin{definition}[Soundness] \label{def:soundness}
An accumulator $\Sigma = (\Acc, \Ver)$ is \emph{sound} if it is infeasible to find a verifying membership proof for an element not included in an honestly committed set. More precisely, there exists a negligible function $\mathsf{negl}$ such that for every computationally unbounded adversary $(\adv_1, \adv_2)$ making polynomially many random oracle queries and every polynomially bounded set size $n = n(\theta, \secpar)$:
\[
\Pr\left[
\begin{array}{l|l}
    \Ver^{\calR}(c,u,\pi)=\true \land u\notin S
    & (S,\mathsf{st}) \gets \adv_1^{\calR}(1^\secpar) \\
    & (c,(\pi_1,\ldots,\pi_n)) \gets \Acc^{\calR}(S) \\
    & (u,\pi) \gets \adv_2^{\calR}(\mathsf{st},c,(\pi_1,\ldots,\pi_n))
\end{array}
\right] \leq \negl.
\]
\end{definition}

\begin{definition}[Correctness] \label{def:correctness}
An accumulator $\Sigma = (\Acc, \Ver)$ is \emph{correct} if an up-to-date membership proof $\pi$ corresponding to an element $u$ in accumulator $c$ can always be used to verify the membership of $u$ in $A$. More precisely, for all security parameter $\secpar$, any polynomially sized subset $S\subseteq \calU$ and all elements $u \in S$:
\[
\Pr_{\calR} \left[
\begin{array}{cc}
     & \Ver^{\calR}(c, u, \pi) = \false: \\
     & c, (\pi_1,\ldots,\pi_{n}) \gets \Acc^{\calR}(S)
\end{array} \right] \leq \negl,
\]
where $\pi$ is the membership proof corresponding to $u$ with regard to the accumulator $c$.
\end{definition}

\begin{remark}
    As with vector commitments, we consider accumulators with deterministic verifiers.
    Observe that \Cref{fact:verifier-transform} applies to accumulators as well, allowing us to assume without loss of generality that witnesses consist of tuples of oracle queries.
\end{remark}

\section{The Communication Protocol}
In this section, we present the proof of our lower bound, assuming the disjoint product lemma, which we state and prove in the next section (\Cref{sec:crabs}).
We focus on the version of the bound for vector commitments but extend it to accumulators as a corollary (\Cref{cor:acc}).

We first define some notions that will be useful in describing the communication protocol (shown in \Cref{alg:communication}).

\begin{definition}[Certified.]
    We say that an index-bit pair $(i, b)$ is \emph{certified} by a table $T \in \left(\zo^{\theta + \secpar} \right)^*$ with respect to a commitment $c$ if
    there exist at most $k$ table entries $W \subseteq T$ such that
    \[
    \Ver^{\calR}(c, i, b, W) = 1
    \]
    for any $\calR$ consistent with $T$.
\end{definition}
We also say an index $i$ alone is certified if there exists a bit $b$ such that $(i, b)$ is certified. We next define residual query set.

\begin{definition}[$p$-hard.]
    We say that a query-answer table $T$ is \emph{$p$-hard} with respect to a commitment $c$ if every $p$-query algorithm has probability at most $1/p$ of finding and querying an accepting witness for any uncertified index under $T$.
    This probability is over a random oracle extending $T$.
\end{definition}

\begin{definition}[Residual query set.]
    Let $\calR$ be a random oracle, $T$ be a table, and $x \in \zo^n$ be a vector.
    Let $c, \pi_1, \ldots, \pi_n \gets \Com^{\calR}(x)$, and for each $i \in [n]$ let $W_i$ denote the set of random oracle inputs in $\pi_i$.
    The \emph{residual query set} of an index $i \in [n]$ is $W_i \setminus \dom(T)$, the honest witness' queries that are missing from $T$.
\end{definition}
Our proof proceeds via a compression argument. Before describing the compression protocol, we introduce and analyze the procedure $\Norm_p$ (\Cref{alg:normalize}) that, given a commitment $c$ and a table $T$ of random-oracle query-answer pairs, attempts to certify indices $i \in [n]$.

\begin{algorithm}[t!]
\caption{$\Norm_p(c,T)$, using hardness parameter $p$.}\label{alg:normalize}
\noindent\begin{minipage}{\dimexpr\linewidth-34pt\relax}
\vspace{8pt}
\textbf{Input:} oracle access to $\calR$; a commitment $c$ and a partial table $T$.\\
\textbf{Output:} a table extending $T$, and a status in $\{\st{binding error},\st{hard},\st{fail}\}$.\\
For $j=1,2,\dots,3p n^2\secpar + 1$:
\begin{enumerate}[label=\arabic*.,ref=\arabic*,leftmargin=2em]
\item If $T$ certifies both $b = 0$ and $b = 1$ for the same position, return $(T,\st{binding error})$. If $T$ is $p$-hard, return $(T,\st{hard})$. If $j=3pn^2\secpar +1$, return $(T,\st{fail})$.
\item\label{st:tree} Let $A$ be the first deterministic $p$-query decision tree that outputs an accepting witness for any uncertified index under $T$ with probability greater than $1/p$, when run on a random oracle extending $T$.
\item Run $A$ on $\calR$. For each random oracle query $q \in \zo^\theta$ made by $A$, append $(q, \calR(q))$ to $T$. If $A$ outputs an alleged witness $W_i$, make these queries as well and add them to $T$.
\end{enumerate}
\end{minipage}
\vspace{8pt}
\end{algorithm}
We prove the following properties of the $\Norm_p$ procedure.
\begin{lemma}[Hardening]\label{lem:normalize}
The procedure $\Norm_p(c,T)$ has the following properties.
\begin{enumerate}[label=(\alph*)]
\item It makes at most $(3pn^2\secpar + 1)(p + k) \leq 4p^2 n^2\secpar$ oracle queries.
\item It is a deterministic oracle procedure with input $(c,T)$. Its returned table is the starting table together with all answers it received, and it is hard if the status is \st{hard}.
\item Let $c$ and a table $T_0$ be fixed, and let $\calR$ be uniform among the functions extending $T_0$. Then $\Norm_p(c,T_0)$ returns \st{fail} with probability at most $2^{-n\secpar}$.
\end{enumerate}
\end{lemma}

\begin{proof}
    For (a), observe that in each round we make at most $p$ queries to run the decision tree, plus $k$ queries for the returned alleged witness. This makes $(3pn^2 \secpar + 1)(p + k)$ queries across all rounds.

    For (b), observe that the procedure is deterministic and returns \st{hard} only if $T$ is $p$-hard by construction.

    We prove (c) by a simple probability bound. For $j \in [3pn^2\secpar + 1]$, let $X_j$ be a Bernoulli random variable that is 1 if and only if the algorithm certifies a new in round $j$ of $\Norm$.
    Each $X_j$ has expected value at least $1/p$ conditioned on $X_{< j}$.

    If $\Norm$ terminates early, we define $X_j = 1$ for all subsequent $j$.

    Observe that $\Norm$ returns \st{fail} only if it fails to certify a new index in all but at most $n$ rounds:
    \[
    \Sigma_{j \in [3pn^2\secpar + 1]} X_j < n.
    \]

    By a counting argument, if this occurs, there must be a contiguous interval of length $3pn\secpar$ such that all $X_j$'s in this interval are 0.
    For a fixed interval, this occurs with probability at most
    \[
    \left(1 - 1/p \right)^{3pn \secpar} \leq e^{-3n\secpar}.
    \]
    By a union bound, there exists such an interval with probability at most $(3pn^2\secpar + 1)e^{-3n\secpar} \leq 2^{-n\secpar}$.
\end{proof}
Our compression protocol is given in \Cref{alg:communication}. We are now ready to analyze it. We next show that the probability the protocol terminates with an error is negligible. We start by bounding the probability of a hardening error.

\begin{algorithm}[t!]
\caption{Communication protocol between Alice and Bob. }
\label{alg:communication}
\vspace{8pt}

\noindent\begin{minipage}{\dimexpr\linewidth-34pt\relax}

\textbf{Setup (free communication).}
\begin{itemize}[leftmargin=1.5em, itemsep=7pt, topsep=4pt]
    \item
    Let $Q$ be an upper bound on the number of random oracle queries made by an honest committer to an $n$-bit vector. Set $p := Q + 2kn + n^3$.
\end{itemize}

\medskip
\textbf{Initialization.}
\begin{itemize}[leftmargin=1.5em, itemsep=7pt, topsep=4pt]
    \item \textbf{Bob:} Initialize $T\gets\emptyset$.
    \item \textbf{Alice:} Sample $x \gets \{0,1\}^n$.
    Compute $(c,\pi_1,\ldots,\pi_n)
    \gets \Com^{\calR}(x)$. If any witness fails to verify (i.e., $\Ver^{\calR}(c, i, x_i, \pi_i) = \false$ for some $i$), \textbf{terminate with completeness error}.

    Send $\alpha_0 := c$ to Bob.
\end{itemize}

\medskip
\textbf{For $i=1,\ldots,k$:}
\begin{itemize}[leftmargin=1.5em, itemsep=7pt, topsep=4pt]
    \item \textbf{Bob (hardening):}
    Run $\Norm_p(c,T)$. If the status is \st{fail}, \textbf{terminate with hardening error}.
    If the status is \st{binding error}, \textbf{terminate with binding error}.
    \item \textbf{Alice:}
    Alice follows Bob's execution, as Bob is deterministic, and maintains a consistent copy of $T$.
    Let $S = \{s_1, \ldots, s_m\} \subseteq [n]$ be the set of uncertified indices under $T$ and let $W_j$ be the residual query set of $s_j$.
    Alice greedily finds maximal subset $S^* \subseteq S$ whose residual sets are pairwise disjoint.
    If $\abs{S^*} \geq \ell_{i-1} + \secpar$, where $\ell_{i-1}$ is the total number of bits Alice has sent through the end of round $(i-1)$, \textbf{terminate with disjointness error}.

    \item \textbf{Alice $\longrightarrow$ Bob:}
    Send the query set
    \[
    \alpha_i = \bigcup_{s_j \in S^*} W_j,
    \]
    padded to bit length $2k \theta (\ell_{i-1} + \secpar)$.
    Alice checks if adding the queries in $\alpha_i$ to $T$ certifies any index $j$ for the wrong bit $x_j$. If so, we \textbf{terminate with binding error}.

    \item \textbf{Bob (termination check):}
    Query $\calR$ on every input in $\alpha_i$
    and add the query-answer pairs to $T$.
    If any $j \in [n]$ is certified under $T$ for both $(j, 0)$ and $(j, 1)$, \textbf{terminate with binding error}.
    If all $j \in [n]$ are certified for exactly one bit, \textbf{terminate with success and output} the vector of certified bits.
\end{itemize}

\end{minipage}
\vspace{8pt}
\end{algorithm}

\begin{lemma}[Negligible probability of hardening error] \label{lem:hard}
If Alice sends at most $n$ bits, the probability that $\Norm$ terminates in $\mathsf{fail}$ is negligible.
\end{lemma}

\begin{proof}
    We already showed that $\Norm$ terminates with $\mathsf{fail}$ with probability at most $2^{-n\secpar}$ when run on a fixed commitment $c$, table $T_0$, and random oracle $\calR$ that is \emph{uniform among those extending} $T_0$.
    However, in the real protocol, the random oracle is correlated with the commitment and Alice's message.

    Instead, we bound
    \begin{align*}
    \Pr_{\calR}[\calR\text{ produces }c, T_0 \text{ that induce }\mathsf{fail} \gets \Norm_p(c, T_0)]
    \end{align*}
    $T_0$ is fully determined by Alice's transcript $\alpha$, which includes $c$, and Bob's oracle answers.
    By assumption, the length of $\alpha$ is upper bounded by $n$.
    Therefore, we can apply a union bound:
    \begin{align*}
        \Pr_{\calR}[\calR\text{ produces }\alpha \text{ that induces }\mathsf{fail} \gets \Norm_p(c, T_0)] &= \sum_{\alpha \in \zo^{\leq n}}\Pr_{\calR}[\calR\text{ produces }\alpha \land \mathsf{fail}]\\
        &\leq \sum_{\alpha \in \zo^{\leq n}} \Pr_{\calR}[\mathsf{fail}]\\
        &\leq 2^{n + 1} \cdot 2^{-n\secpar},
    \end{align*}
    which is negligible.

    Note that we did not condition on the event that Alice sends $\alpha$; rather, we bounded the probability that Alice sends $\alpha$ \emph{and} the bad event occurs.
    We will use this same argument to avoid conditioning again in the next proof.
\end{proof}

We next use the disjoint product lemma (\Cref{lem:crab}) to bound the probability of a disjointness error.

\begin{lemma}[Negligible probability of disjointness error] \label{lem:disj-error}
    If $k \leq \log n$, the probability that the communication protocol terminates with disjointness error is negligible.
\end{lemma}
\begin{proof}

    Fix a commitment $c$ and $p$-hard table $T$.
    Let $\calX = \zo^\theta \setminus \dom(T)$. Let $S$ be the set of all uncertified indices under $T$, and let $\calR_T$ be a random oracle from $\calX \to \zo^\secpar$. Define the following local witness predicate on
    nonempty sets $W\subseteq\calX$ with $|W|\le k$:
    \[
    V^{\calR_T}_{S,c,T}(W)=1
    \]
    if and only if there exist $j\in S$ and $b\in\zo$
    such that $\Ver(c,j,b,\pi)$ accepts for some opening
    $\pi$, whose queries lie only in
    $\dom(T)\cup W$.

    Hardness of the table implies that the success probability of any $p$-query algorithm relative to $V$ is at most $1/p$, over the random oracle $\calR_T$ with domain $\calX$. By the disjoint product lemma (\Cref{lem:crab}), any $(p - tk)$-query algorithm $\alg^{\calR_T}$ finds $t$ disjoint verifying sets $W_1, \ldots, W_t \in \calX$ with probability at most $(e^k/p)^t$. Since $k \leq \log n$, this is at most $2^{-t}$.
    In particular, this bounds the probability of certifying any subset of $t$ indices with disjoint residual query sets given $T$.

    We now wish to show that the probability that the honest commitment algorithm $\Com$ finds $t$ witnesses with these disjoint residual sets is also small.
    We cannot apply the disjoint product lemma directly: it considers a random oracle that is independent of the fixed commitment $c$.
    By contrast, the honest committer chooses $c$ depending on its oracle queries.

    Instead, fix in advance a round $i \in [k]$ and an arbitrary partial transcript $\alpha = (c, \alpha_1, \ldots, \alpha_{i-1})$ up to that round.
    Let $\Bad_i(\alpha)$ be the event that Bob's hardening in round $i$ succeeds and, relative to his resulting table $T$, there are at least $t$ uncertified indices whose honest residual query sets are pairwise disjoint.
    Note that all random oracle responses outside of $T$ are uniform and independent of $\alpha$.

    Now consider an algorithm $\alg^{\calR_T, T}$ that runs the honest commitment algorithm on a random $x$, checks if the resulting commitment equals $c$, and outputs $t$ pairwise-disjoint residual query sets if they exist.

    If Alice produces transcript $\alpha$ and $\Bad_i(\alpha)$ holds, $\alg$ succeeds.
    Indeed:
    \begin{itemize}
        \item $T$ is $p$-hard, and in particular finding any new witness is hard under the random oracle $\calR_T$ extending $T$.
        \item Running $\Com^{\calR_T, T}(x)$ on Alice's chosen vector $x$ yields $t$ witnesses that verify under $c$, whose residual sets under $T$ are pairwise disjoint.
        That is, $\Com$ finds $t$ pairwise disjoint solutions to the verifier $V$.
    \end{itemize}
    $\alg$ makes at most $Q \leq (p - tk)$ queries, so our earlier application of the disjoint product lemma shows that this occurs with probability at most $2^{-t}$.

    Therefore, recalling that $t = \ell_{i-1} + \secpar$:
    \begin{align*}
        \Pr[\text{disjointness error in round } i] &\leq \sum_{\alpha \in \zo^{{\ell_{i-1}}}} \Pr[\text{Alice produces transcript }\alpha \land \Bad_i(\alpha)]\\
        &\leq 2^{\ell_{i-1}} 2^{-t}\\
        &= 2^{-\secpar}.
    \end{align*}

     By a union bound, the probability that Alice terminates with a disjointness error in any round is $k \cdot 2^{-\secpar} = \negl$.
\end{proof}

Next, we observe that the binding property of the VC implies that the probability of a binding error is negligible.
\begin{lemma}[Negligible probability of binding error.] \label{lem:bind-error}
The probability that the communication protocol terminates with binding error is negligible.
\end{lemma}

\begin{proof}
    First, observe that in every instance where the protocol terminates with \textbf{binding error}, the queries made by Alice and Bob thus far indeed certify some incorrect bit $\bar{x_j}$:
    \begin{itemize}
        \item A binding error in Bob's hardening step occurs exactly if Bob's table certifies both $b = 0$ and $b = 1$ for some index.
        \item A binding error occurs in the step where Alice sends to Bob exactly if $T$ certifies some index for the wrong bit.
        \item Bob's termination check terminates with binding error if Bob's table certifies both $b = 0$ and $b = 1$ for some index.
    \end{itemize}

    We will now argue that Alice and Bob together make a polynomial number of queries to the random oracle.
    Then, if the communication protocol finds a binding error, this gives a polynomial-query algorithm that breaks position binding for a random vector $x$.

    The protocol makes queries in the following places:
    \begin{itemize}
        \item Alice commits to $x$, which costs at most $Q$ queries.
        \item Alice checks the witnesses she generated, which costs at most $kn$ queries.
        \item In each round $i \in [k]$:
        \begin{itemize}
            \item Bob runs $\Norm$. This costs at most $4p^2 n^2\secpar$ oracle queries.
            \item Alice and Bob both query the set $\alpha_i$.
            Since Alice only sends queries from honest witnesses, she sends at most $kn$ queries over all rounds.
            \item Alice follows Bob's execution. She only ever makes queries that Bob has already made. Therefore, this costs no additional queries for an algorithm simulating Alice and Bob together.
        \end{itemize}
    \end{itemize}

    Therefore, the total number of queries is at most
    \[
    Q + kn + k(4 p^2 n^2\secpar + 2k(kn\theta + \abs{c} + \secpar)).
    \]
    All of these quantities are polynomial in $\secpar$; therefore, an algorithm simulating both Alice and Bob in the protocol makes at most a polynomial number of queries.
\end{proof}
Finally, we bound the probability of a completeness error.
\begin{lemma}[Negligible probability of completeness error.] \label{lem:compl-error}
    The probability that the communication protocol terminates with completeness error is negligible.
\end{lemma}

\begin{proof}
    Completeness implies that for \emph{all} $x \in \zo^n$,
    \[
        \Pr_{\calR}\left[\exists i \in [n] \text{ s.t. } \Ver(c, i, x_i, \pi_i) = \false: \ c, \pi_1, \ldots, \pi_n \gets \Com(x) \right] \leq \negl.
    \]
    Therefore, this probability is also negligible for Alice's random $x \gets \zo^n$.
\end{proof}
\subsection{Proving the Main Theorem}
We are now ready to state and prove our main theorem.
\begin{theorem} \label{thm:main}
    Let $\Sigma = (\Com, \Ver)$ be a vector commitment with a deterministic verifier secure under a random oracle $\calR: \zo^\theta \to \zo^\secpar$, with commitments of bit length $\abs{c}$.
    Then the verifier query complexity $k$ for an $n$-bit vector for $n = \Omega(\secpar)$ must satisfy
    \[
    2 + (3k \theta)^k(\abs{c} + \secpar) \geq n.
    \]
\end{theorem}

\begin{proof}
Assume for the sake of contradiction that $2 + (3k \theta)^k(\abs{c} + \secpar) < n$.
We will show that the communication protocol allows Alice to transmit a random $n$-bit vector to Bob by sending fewer than $n$ bits.
Our proof will proceed as follows:
\begin{itemize}
    \item We first show that the protocol terminates with an error with only negligible probability. This invokes the lemmas just proven.
    \item We show that if the protocol \emph{does} finish without termination, Bob always recovers $x$ exactly.
    \item We prove an upper bound on the number of bits that Alice sends.
    \item We conclude that the protocol allows Alice to communicate beyond information theoretic limits.
\end{itemize}

\textbf{Overall probability of termination in error.}
We've shown that the protocol terminates with each of completeness error (\Cref{lem:compl-error}), hardening error (\Cref{lem:hard}), binding error (\Cref{lem:bind-error}), and disjointness error (\Cref{lem:disj-error})\footnote{Note that our assumption on $k$ implies that $k \leq \log n$, satisfying the assumption in \Cref{lem:disj-error}.} with negligible probability.
Therefore, by a union bound, it terminates with any kind of error with negligible probability.

\paragraph{Correct decoding if the protocol does not terminate with error.}
Suppose that the protocol does not terminate with an error.
We'll first argue that at the end of round $k$, all $n$ indices must be certified.
We start by showing the following invariant:

\begin{fact}
    Suppose the protocol does not terminate.
    Then for all $i \in [k]$, after round $i$ of the protocol, for every index $j$ either:
    \begin{itemize}
        \item $j$ is certified, or
        \item at least $i$ queries of $\pi_j$ are in $T$.
    \end{itemize}
\end{fact}

\begin{proof}
    We prove this by induction.
    The base case, $i = 0$, holds vacuously: After the $0^{\text{th}}$ initialization round, no indices are certified and at least 0 queries of each witness are in the empty table $T$.

    Assume for $i \in [k]$ that after round $(i-1)$ every index is either certified or has at least $(i-1)$ of its honest witness queries in $T$.
    Now, in round $i$, Bob hardens his table. This only adds to $T$.

    The key step is Alice's generation of $\alpha_i$.
    Recall she finds a maximal subset $S^*$ of the uncertified set $S$ whose residual sets are pairwise disjoint. She takes $\alpha_i$ to be the union of the residual sets of indices in $S^*$.
    Since $S^*$ is maximal, every $s_j \in S \setminus S^*$ has at least one residual query in $\alpha_i$; otherwise, Alice could add $s_j$ to $S^*$, contradicting maximality.

    Bob then adds $\alpha_i$ to $T$, which certifies all of $S^*$ and adds at least one residual query for every $s_j \in S \setminus S^*$.
    Therefore, at the end of round $i$ every uncertified index has at least $i$ of its honest witness' queries in $T$.
\end{proof}

This means that after round $k$ of the protocol, every index is either certified or has $k$ queries of its witness in $T$.
In the latter case, the index must also be certified, as witnesses have $k$ queries.

It remains only to verify that no index is certified for the incorrect bit.
Recall that after Alice sends her final query set to Bob, she checks whether the resulting table certifies any incorrect bit and terminates if to.
Therefore, if the protocol does not terminate with error, all indices are certified only for their correct bit.
Bob thus outputs exactly $x$.

\paragraph{Total number of bits sent.}
Recall that $\ell_{k}$ is the total number of bits Alice has sent through the end of round $k$.

\begin{fact}
    The total number of bits Alice sends throughout the protocol is at most $(3k\theta)^k(\abs{c} + \secpar)$.
\end{fact}

\begin{proof}
By construction of the protocol, we have $\ell_0 = \abs{c}$, and for $i \geq 1$ we have
$\ell_{i} = (2k\theta + 1)\ell_{i-1} + 2k \theta \secpar$, and therefore
\[
\ell_i + \secpar = (2k\theta + 1)(\ell_{i-1} + \secpar).
\]
Therefore, we have

\begin{align*}
    \ell_k &= (2k\theta + 1)^k(\abs{c} + \secpar) - \secpar\\
    &\leq (3k\theta)^k (\abs{c} + \secpar).
\end{align*}
\end{proof}

\paragraph{An information theoretic contradiction.}
We've shown that Alice correctly transmits $x$ in $\ell_k$ bits except with negligible failure probability.
She can increase correctness to probability 1 as follows.
After she generates her commitment, and before she sends anything to Bob, she simulates the entire protocol.
If the protocol terminates with error, she sends `0' followed by $x$.
If the protocol is successful, she sends `1' and proceeds with the protocol.

If Bob receives `0' as his first bit, he outputs the following $x$. Otherwise, he proceeds with the protocol.

The protocol now succeeds in transmitting a random $n$-bit vector with probability 1, and the expected number of bits Alice transmits is
\[
1 + (1-p(\text{fail})) \cdot \ell_k + p(\text{fail}) \cdot n.
\]
Since $p(\text{fail})$ is negligible, this is at most $2 + \ell_k \leq 2 + (3k\theta)^k (\abs{c} + \secpar)$.
By assumption, this is less than $n$, yielding our information theoretic contradiction.
\end{proof}
\subsection{Corollaries}
We derive two corollaries on the number of queries of vector commitments and accumulators, and one on the bit complexity of vector commitment verifiers.
\begin{corollary}[of \Cref{thm:main}] \label{cor:bound}
    Consider any vector commitment with a deterministic verifier that is secure in the random oracle model.
    For any commitment size $\abs{c} = O(n^\epsilon)$ for some constant $\epsilon < 1$, for sufficiently large $n$ we have
    \[
    k = \Omega(\log_\theta n).
    \]
\end{corollary}

\begin{proof}
    Suppose that $k \leq \log n$ (and observe that if not, the corollary already holds). We expand the bound obtained in \Cref{thm:main}, first rewriting:

    \begin{align*}
        k  &\geq \frac{\log \frac{n-2}{\abs{c} + \secpar}}{\log(3k\theta)}\\
        &\geq \frac{\log \frac{n-2}{\abs{c} + \secpar}}{\log(3\theta \log n)}, \text{ since } k \leq \log n\\
        &\geq \Omega \left(\frac{\log n}{\log(\theta) + \log \log n}\right), \text{ since }n/(\abs{c} + \secpar) = \Omega(n^{1 - \epsilon - \delta})\text{ for some constant } \delta > 0\\
        &\geq \Omega \left(\frac{\log n}{\log \theta} \right) \text{ since $n = o(2^\theta)$}.
    \end{align*}

\end{proof}

\begin{corollary}[of \Cref{thm:main}] \label{cor:acc}
    Consider any \emph{accumulator} with a deterministic verifier that is secure in the random oracle model.
    For any commitment size $\abs{c} = O(n^\epsilon)$ for some constant $\epsilon < 1$, for sufficiently large $n$ we have
    \[
    k = \Omega(\log_\theta n).
    \]
\end{corollary}

\begin{proof}
We briefly adapt the communication protocol and invocations of position binding to cryptographic accumulators and their slightly different security notion.

First, we adapt the way that Alice commits to her vector, as accumulators commit only to sets.
For $x \in \zo^n$, Alice uses the data universe $[n] \times \zo$ and accumulates the set
\[
S = \{(i, x_i)\}_{i \in [n]}.
\]
The effective witness for component $i$ of Alice's vector is the witness for $(i, x_i)$.
Correctness of the accumulator is analogous to completeness of the vector commitment: honestly generated witnesses verify except with negligible probability.

Accumulator security states that if $c$ is an honestly generated commitment for an adversarially chosen set $S$, it is infeasible to produce an element $u \notin S$ and a verifying proof with respect to $c$.
The commitment is indeed honestly generated in our protocol.
It remains to check that a contradiction to position binding implies a contradiction to accumulator security.

Recall that we arrive at a position binding error if either Alice or Bob finds verifying proofs for both $x_i = 0$ and $x_i = 1$.
That is, they find proofs for both $(i, 0)$ and $(i, 1)$.
However, Alice included only one of these tuples in her set; the proof for the excluded tuple contradicts accumulator security.
\end{proof}

We now observe that our lower bound can be immediately extended to show a tradeoff between the verifier's query complexity and the bit length of the witnesses.
This bound is informative only for small verifier complexity $k$ and is not tight; we present it primarily as guidance for future work.

\begin{corollary}[of \Cref{thm:main}] \label{cor:query-communication}
    Let $\Sigma = (\Com, \Ver)$ be a vector commitment with a deterministic verifier secure under a random oracle $\calR$, with commitments of bit length $\abs{c}$ and witnesses of bit length $\abs{\pi_i}=w$.
    Then the verifier query complexity $k$ for an $n$-bit vector  for $n = \Omega(\secpar)$ must satisfy

    \[ w \ge \Omega\left(\left(\frac{n}{|c|+\secpar}\right)^{1/k}\right) -  \ceil{\log n} . \]
\end{corollary}

Note that this bound is informative only when $k = o(\log n)$.
In the regime of high-arity Merkle trees, where $k = \log_{(\secpar/\log^2 \secpar)} n$ and $\abs{c} = \Theta(\secpar)$, it is tight up to a factor $(\secpar^{1/k} \cdot \log n \cdot \log \secpar)$.
That is, the lower bound gives $w \geq \Omega(\secpar^{1 - 1/k}/\log^2 \secpar)$.
The witness length achieved by a $(\secpar/\log^2 \secpar)$-ary Merkle tree is $O(\secpar \cdot \log_{\secpar}(n))$.

\begin{proof} We explain how to adjust the proof of \Cref{thm:main} to derive that
    \[
    2 + (3(w+\ceil{\log n}))^k(\abs{c} + \secpar) \geq n.
    \]

    \Cref{cor:query-communication} then follows by rearranging the expression.
    The main observation is that to describe a residual set $W_j$, instead of describing the queries explicitly using  $k\theta$ bits Alice can send the witness $\pi_j$, and Bob can execute $\Ver$ to retrieve its queries. We make the following change to the proof.

    In the communication protocol (\Cref{alg:communication}), to describe the query set $\alpha_i$, Alice sends $(j,x_j,\pi_j)$ for each $j\in S^*$, padded to length $2(w+\ceil{\log n})(\ell_{i-1}+\secpar)$ (note that for each $j$, $(j,x_j,\pi_j)$ can be described using $(w+\ceil{\log n}+1)$ bits). To retrieve $W_j$, Bob simply executes $\Ver^\R(c,j,x_j,\pi_j)$ and adds all of its query-answer pairs to $T$. The proof now continues similarly to the proof of  \Cref{thm:main}.

\end{proof}

\section{Proof of the Disjoint Product Lemma} \label{sec:crabs}

We now turn to proving the disjoint product lemma that underlies our proof.

\begin{lemma}[Disjoint Product Lemma]
\label{lem:crab}
    Let $\calX$ be a set, $\calR: \roc$ be a random oracle, and $V^{\calR}$ be a deterministic verifier, which takes as input at most $k$ points $q_1, \dots, q_{k'} \in \calX$, queries $\calR(q_1), \dots, \calR(q_{k'})$ and then accepts or rejects. Let $t$ be an integer and suppose that any $(Q + tk)$-query algorithm finds a verifying set to the verifier with probability at most $\eta$. Then any $Q$-query algorithm computes $t$ disjoint sets $W_1, \dots, W_t$, which are all accepted by the verifier, with probability at most $e^{tk} \cdot \eta^t$.
\end{lemma}

Roughly, the lemma says that if it is hard to find a single verified solution then it is exponentially harder to find $t$ disjoint ones.\footnote{A knowledgeable reader may be reminded of crab mentality, the notion that ``if I can't have it, nobody can.'' Indeed, if a single algorithm cannot compute a solution, then it will bring down the whole bunch of $t$ solutions. For this reason, the authors affectionately refer to this lemma as the ``crab'' lemma, but have avoided doing so in this write-up to ease presentation to a less lettered audience. \label{foot:crab}} We remark that the proof was found by OpenAI's Astra model and refer the reader to our statement on AI usage for more details.

To prove the statement, we show that if there exists a $Q$-query algorithm $\alg$ that finds $t$ disjoint verifying sets with probability $\delta$, then there is a $(Q+tk)$-query algorithm that finds a single verifying set with probability at least $e^{-k} \delta^{1/t}$. Using the premise of \Cref{lem:crab}, we then get that $e^{-k} \delta^{1/t} \leq \eta$ and rearranging yields the result. Before discussing how we construct such an algorithm from $\alg$, we describe three simplifying assumptions that we make throughout this section.

First, we will assume that $\alg$ always queries the values in $W_1, \dots, W_t$. Under this assumption, it then suffices to prove that there is a $Q$-query algorithm that finds a single verifying set with probability at least $e^{-k} \delta^{1/t}$, as we can simply query all points in $\bigcup_{i=1}^t W_i$ at the end at the cost of an additional $tk$ queries.

Second, we will assume that $\alg$ is deterministic. Notably, if there exists a randomized algorithm $\alg$ that computes $t$ disjoint verifying sets with probability at least $\delta$, then there must be some seed for which $\alg$ computes such disjoint sets with probability at least $\delta$. Given this, we can assume $\alg$ is deterministic and thus corresponds to a decision tree, $\calT$, of depth at most $Q$. Each internal vertex, $v$, in the tree is labeled by a query $\ell(v) \in \calX$ and we branch based on the $2^{\secpar}$ possible values of the random oracle value $\calR(\ell(v))$, yielding the children $vy$ for $y \in \zo^{\secpar}$. We will write $\emptyset$ as the root and let vertices be described by the answers from the oracle to reach them in the tree, i.e. they are of the form $y_1 y_2 y_3 \dots y_i$ with $y_i \in \zo^{\secpar}$.

Third, we will assume that $\alg$ never repeats a query along a root-to-leaf path of $\calT$, as it can simply remember the answers to its earlier queries. Thus, the answers of a random oracle along any root-to-leaf path are independent and uniform; in particular, a leaf at depth $m$ is reached with probability $2^{-\secpar m}$.

With these assumptions, we now describe how we find a single verifying set from $\calT$. In particular, this will be done via a simulation argument. In an ideal world we could simply follow $\calT$ conditioned on reaching a ``good leaf,'' i.e. those leaves in the decision tree with $t$ disjoint verifying sets to the verifier. Unfortunately, it's easy to see that such a scenario is too much to hope for as a random oracle only reaches a good leaf with probability $\delta$. As such, we expect following the random oracle to not correspond to our ideal execution.

That said, we do have some leeway in the path that we follow down the tree. In particular, we expect to make $Q \gg k$ queries along a given path and whether a set $T$ is accepted by the verifier $V$ only requires the queries in $T$ to be answered correctly by the oracle. Accordingly, we will not mandate that we follow the path in $\calT$ dictated by the random oracle. Instead, we create a new oracle $\ideal$, corresponding to a random oracle conditioned on always reaching a good leaf. We then couple $\ideal$ and the true random oracle $\calR$ to maximize the probability that they agree, follow the path according to $\ideal$, and output a random set among $W_1, \dots, W_t$.

\subsection{The Ideal Oracle and Coupling It to Reality}
We begin by working out a more explicit formulation of the ideal oracle, $\ideal$. Notably, $\ideal$ will not be distributed according to a product distribution and the output of $\ideal$ on a point will be dependent on the node $v$ that we are at in the decision tree. In particular, querying $\ideal$ at node $v$ on $\ell(v)$ in the tree yields the answer $y$ with probability\cal
\[\mu_v(y) := \frac{ \Pr_{\calR} \left[\text{we end at a good leaf } \bigwedge \text{ we reach $vy$} \right] }{\Pr_{\calR} \left[\text{we end at a good leaf } \bigwedge \text{ we reach $v$} \right]}.\]

Now let $\tau = y_1 \dots y_{m}$ be a good leaf. Recall that this notation means that we reached $\tau$ by starting at the root of $\calT$ and receiving the answers $y_1, y_2, \dots, y_{m}$ in that order in our queries to the oracle, $\calR(\ell(\emptyset)), \calR(\ell(y_1)), \calR(\ell(y_1 y_2)), \dots, \calR(\ell(y_1 \dots y_{m-1}))$.
We can then verify that under $\ideal$, we reach $\tau$ with probability
\begin{align*}
    \prod_{i = 1}^{m} \frac{ \Pr_{\calR} \left[\text{we end at a good leaf } \bigwedge \text{ we reach $y_1 \dots y_{i}$} \right] }{\Pr_{\calR} \left[\text{we end at a good leaf } \bigwedge \text{ we reach $y_1 \dots y_{i-1}$} \right]} &= \frac{\Pr_{\calR} \left[ \text{we reach $\tau$} \right]}{\delta} \\
    &= \Pr_{\calR} \left[ \text{we reach $\tau$} \bigg| \text{we end at a good leaf} \right],
\end{align*}

as desired.

Now as mentioned, we wish to couple $\ideal$ and the true random oracle to maximize their probability of agreement. To do so, we will simply use a greedy coupling. Suppose that we have reached node $v$ in the tree and let $y = \calR(\ell(v))$ be the true random oracle value. If $\mu_v(y) \geq 2^{-\secpar}$, we set $\ideal(\ell(v)) = \calR(\ell(v))$. If, on the other hand, $\mu_v(y) \leq 2^{-\secpar}$, then we sample a Bernoulli random variable with mean $2^{\secpar} \mu_v(y)$. If this variable is $1$, then we set $\ideal(\ell(v)) = y$. If the Bernoulli variable is $0$, then we set $\ideal(\ell(v)) = z$ with probability proportional to $(\mu_v(z) - 2^{-\secpar})_+$, where $(x)_+ := \max\{x, 0\}$.

It will then be useful to understand the probability, under this coupling, that a set of queries along a root-to-leaf path in $\calT$ agrees under $\ideal$ and $\calR$.

\begin{lemma}
\label{lem:path-coupling-agreement}
    Let $\tau$ be a good leaf. Denote by $v_1, \dots, v_s$ arbitrary distinct internal vertices %
    along the path from the root of $\calT$ to $\tau$ and let $y_i$  be the values of $\ideal(\ell(v_i))$ corresponding to the path to $\tau$, then
        \[\Pr_{\calR, \ideal} \left[ \calR(\ell(v_i)) = y_i \quad \forall i \in [s] \bigg | \text{Reach $\tau$ under $\ideal$} \right] = \prod_{i=1}^s \min \left( \frac{1}{\mu_{v_i}(y_i)} \cdot 2^{-\secpar}, 1 \right).\]
\end{lemma}

\begin{proof}
    Note that we can equivalently view our coupling as defining $\calR$ after sampling $\ideal$. In particular, at a node $v$ we first sample $y = \ideal(\ell(v))$. If $\mu_v(y) \leq 2^{-\secpar}$, then we set $\calR(\ell(v)) = y$. Otherwise, sample a Bernoulli random variable with mean $\frac{1}{\mu_v(y)} \cdot 2^{-\secpar}$. If this variable is $1$, set $\calR(\ell(v)) = y$. If it's $0$, then set $\calR(\ell(v)) = z$ with probability proportional to $(2^{-\secpar} - \mu_v(z))_+$.

    We can then see that

    \begin{align*}
      \Pr_{\calR, \ideal} \left[ \calR(\ell(v_i)) = y_i \quad \forall i \in [s] \bigg | \text{Reach $\tau$ under $\ideal$} \right] = \prod_{i=1}^s \min \left( \frac{1}{\mu_{v_i}(y_i)} \cdot 2^{-\secpar}, 1 \right),
    \end{align*}

    as we can imagine first sampling all the values of $\ideal$ to reach $\tau$ and then using our independent Bernoulli random variables from the coupling to sample $\calR(\ell(v_i))$.
\end{proof}

\subsection{Analyzing the Algorithm}
We now imagine drawing a random leaf $\tau$ by following $\ideal$. Upon reaching $\tau$, our simulation obtains the $t$ disjoint verifying sets $W_1(\tau), \dots, W_t(\tau)$ chooses one of them uniformly at random to output, which we denote by $T(\tau)$. Since the verifier is deterministic and accepts $T(\tau)$ under $\ideal$, it also accepts $T(\tau)$ under the true oracle $\calR$ whenever $\calR$ and $\ideal$ agree on all the queries in $T(\tau)$. We will show that this happens with probability at least $e^{-k} \delta^{1/t}$.

Towards this, fix a good leaf $\tau = y_1 \dots y_m$ and let $v_1, \dots, v_m$ be the internal vertices along the root-to-$\tau$ path, so that $v_i = y_1 \dots y_{i-1}$ and $y_i$ is the answer received at $v_i$. For each $i \in [m]$, define
\[ r_i := \frac{2^{-\secpar}}{\mu_{v_i}(y_i)}, \]
which gives the ratio between the probability that a random oracle answers $y_i$ at $v_i$ and the probability that $\ideal$ does. Note that $r_i$ is well defined since $\mu_{v_i}(y_i) > 0$ along any path that $\ideal$ follows. By \Cref{lem:path-coupling-agreement}, conditioned on reaching $\tau$ under $\ideal$, the oracles $\calR$ and $\ideal$ agree on $\ell(v_i)$ with probability $\min\{1, r_i\}$, independently across $i$.

Since $\alg$ queries every point of its output sets and never repeats a query along a path, each point of $W_j(\tau)$ is queried at exactly one vertex along the root-to-$\tau$ path. We thus define $I_j \subseteq [m]$ to be the set of indices $i$ such that $\ell(v_i) \in W_j(\tau)$, and note that $|I_j| = |W_j(\tau)| \leq k$. Now, by \Cref{lem:path-coupling-agreement}, conditioned on reaching $\tau$ under $\ideal$, the oracles $\calR$ and $\ideal$ agree on all of $W_j(\tau)$ with probability
\[ a_j(\tau) := \prod_{i \in I_j} \min\{1, r_i\}. \]
As our simulation outputs each $W_j(\tau)$ with probability $1/t$, we can then compute
\begin{align}
    \Pr_{\calR, \ideal} \left[\text{$V$ accepts $T(\tau)$} \right] \geq \E_{\tau \sim \ideal} \left[ \frac{1}{t} \sum_{j=1}^t a_j(\tau) \right]. \label{eq:prob-lb}
\end{align}

It now remains to argue that the right-hand side of \Cref{eq:prob-lb} is not too small. Promisingly, we can note that based on how we defined $\ideal$ the product of the ratios along any path to a good leaf is exactly $\delta$.

\begin{lemma}
\label{lem:product-r}
    Let $\tau = y_1 y_2 \dots y_{m}$ be a good leaf, then
        \[\prod_{i=1}^{m} r_i = \delta.\]
\end{lemma}

\begin{proof}
    Recall that we have that
        \[\Pr_{\ideal} \left[\text{we reach $\tau$} \right] = \prod_{i=1}^{m} \mu_{v_i}(y_i).\]
    On the other hand, we also have
        \[\Pr_{\ideal} \left[\text{we reach $\tau$} \right] = \Pr_{\calR} \left[\text{we reach $\tau$} \bigg | \text{we end at a good leaf} \right] = \frac{\Pr_{\calR} \left[\text{we reach $\tau$} \right]}{\delta} = \frac{2^{-\secpar m}}{\delta}.\]
    Combining these two then yields that
        \[\prod_{i=1}^{m} r_i = \frac{2^{-\secpar m}}{\prod_{i=1}^{m} \mu_{v_i}(y_i)} = \frac{2^{-\secpar m}}{2^{-\secpar m} / \delta} = \delta. \qedhere\]
\end{proof}

If every ratio satisfied $r_i \leq 1$, we would now be done. Indeed, by the arithmetic-geometric mean inequality and the disjointness of $I_1, \dots, I_t$, we would have $\frac{1}{t} \sum_{j} a_j(\tau) \geq (\prod_{i \in I_1 \cup \dots \cup I_t} r_i)^{1/t} \geq (\prod_{i=1}^m r_i)^{1/t} = \delta^{1/t}$, as dropping factors that are at most $1$ can only increase the product. Unfortunately, some ratios may exceed $1$, namely when $\ideal$ is less likely than a random oracle to give the answer $y_i$. To handle these, we use that $r e^{-r} \leq \min\{1, r\}$ for every $r > 0$, so that each ratio can be paid for at the cost of a factor $e^{-r_i}$. In expectation, this is not too expensive as each $r_i$ is at most $1$ in expectation at every step. Formally, we first show the following pointwise bound.

\begin{lemma}
\label{lem:pointwise-product}
    For every good leaf $\tau = y_1 \dots y_m$,
        \[\frac{1}{t} \sum_{j=1}^t a_j(\tau) \geq e^{-k} \delta^{1/t} \exp \left( - \frac{1}{t} \sum_{i=1}^{m} (r_i - 1) \right).\]
\end{lemma}

\begin{proof}
    Let $I := \bigcup_{j=1}^t I_j$. Since the sets $W_j(\tau)$ are pairwise disjoint, so are the sets $I_j$. By the arithmetic-geometric mean inequality and the disjointness of the $I_j$, we then have that
    \begin{align}
        \frac{1}{t} \sum_{j=1}^t a_j(\tau) \geq \left( \prod_{j=1}^t a_j(\tau) \right)^{1/t} = \left( \prod_{i \in I} \min\{1, r_i\} \right)^{1/t}. \label{eq:amgm-product}
    \end{align}
    We now lower bound the product on the right-hand side. For every $r > 0$, we have that $\min\{1, r\} \geq r e^{-r}$ and $1 \geq r e^{1 - r}$. Thus, we obtain
    \begin{align*}
        \prod_{i \in I} \min\{1, r_i\} &= \prod_{i \in I} \min\{1, r_i\} \prod_{i \in [m] \setminus I} 1 \\
        &\geq \prod_{i \in I} \left( r_i e^{-r_i} \right) \prod_{i \in [m] \setminus I} \left( r_i e^{1 - r_i} \right) \\
        &= \left( \prod_{i=1}^{m} r_i \right) \exp \left( - |I| - \sum_{i=1}^{m} (r_i - 1) \right) \\
        &\geq \delta e^{-tk} \exp \left( - \sum_{i=1}^{m} (r_i - 1) \right),
    \end{align*}
    where the last line uses \Cref{lem:product-r} and $|I| \leq tk$. Substituting into \Cref{eq:amgm-product} yields the lemma.
\end{proof}

With this we can now prove \Cref{lem:crab}.

\begin{proof}[Proof of \Cref{lem:crab}]
Recall that we wish to show that our simulation procedure for $\alg$ yields a procedure that outputs a verifying set with probability at least $e^{-k} \delta^{1/t}$. Combining \Cref{eq:prob-lb} with \Cref{lem:pointwise-product}, we have that
\begin{align*}
    \Pr_{\calR, \ideal} \left[ \text{$V$ accepts $T(\tau)$} \right] &\geq \E_{\tau \sim \ideal} \left[ \frac{1}{t} \sum_{j=1}^t a_j(\tau) \right] \\
    &\geq e^{-k} \delta^{1/t} \cdot \E_{\tau \sim \ideal} \left[ \exp \left( - \frac{1}{t} \sum_{i=1}^{m} (r_i - 1) \right) \right] \\
    &\geq e^{-k} \delta^{1/t} \cdot \exp \left( - \frac{1}{t} \E_{\tau \sim \ideal} \left[ \sum_{i=1}^{m} (r_i - 1) \right] \right) \\
\end{align*}
where the third line follows by Jensen's inequality. We now note that we can compute the expectation over $\tau \sim \ideal$ by grouping terms by the internal vertices at which they occur to get

\begin{align*}
        \E_{\tau \sim \ideal} \left[ \sum_{i=1}^{m} (r_i - 1) \right] &= \sum_{v} \Pr_{\ideal} \left[ \text{we reach $v$} \right] \sum_{y:\, \mu_v(y) > 0} \mu_v(y) \left( \frac{2^{-\secpar}}{\mu_v(y)} - 1 \right) \\
        &= \sum_{v} \Pr_{\ideal} \left[ \text{we reach $v$} \right] \left( 2^{-\secpar} \cdot \left| \left\{ y : \mu_v(y) > 0 \right\} \right| - 1 \right) \\
        &\leq 0,
    \end{align*}
    where $v$ ranges over the internal vertices of $\calT$ reached with positive probability under $\ideal$, and the last step holds since there are at most $2^{\secpar}$ possible answers $y$. Thus, we conclude that
    \[\Pr_{\calR, \ideal} \left[ \text{$V$ accepts $T(\tau)$} \right] \geq e^{-k} \cdot \delta^{1/t}. \]

Since $\calT$ has depth at most $Q$, it follows that our simulated algorithm makes at most $Q$ queries. Thus, by the hypothesis of \Cref{lem:crab}, we get that $e^{-k} \delta^{1/t} \leq \eta$. Rearranging then gives that $\delta \leq e^{tk} \eta^t$ as desired.
\end{proof}

\section{A Lower Bound for the Disjoint Product Lemma}

In this section, we aim to show our lower bound for finding disjoint witnesses. In particular, we will prove the following lemma.

\begin{lemma}
\label{lem:lb}
For any $\alpha \in (0,1]$ and $t$ sufficiently large, there exist a set $\calX$ and a verifier $V$ that queries a random oracle $\calR: \calX \rightarrow \zo$ in at most $k = \log^{O(1/\alpha)}(|\calX|)$ positions such that
    \[\eta_t \geq \eta_1^t \cdot e^{t k^{1-O(\alpha)}} \]
where $\eta_t$ denotes the maximum probability a $\textsf{poly}(k)$-query algorithm succeeds at finding $t$ disjoint solutions and $\eta_1$ is the maximum probability that a $2^{k^{O(\alpha)}}$-query algorithm succeeds at outputting a single solution.
\end{lemma}

Roughly, this suggests that in general one cannot hope to do much better than the disjoint product lemma without imposing additional structure or constraints on the parameters. The core step in the proof will be the following lemma due to Astra.

\begin{lemma}
\label{lem:raw-lb}
    For any sufficiently large $t$, there exist a set $\calX$ of size $\exp(\textsf{poly}(t))$, a random oracle $\calR: \calX \rightarrow \zo$, and a verifier that makes $O(t^7)$ queries such that $(i)$ there exists an algorithm making $\textsf{poly}(t)$ queries that finds $t$ disjoint solutions with probability at least $\frac{1}{2}$, and $(ii)$ any algorithm that makes $2^{O(t)}$ queries finds a single solution with probability at most $\frac{1}{2} + O(1/t)$.
\end{lemma}

Informally, the lemma roughly yields a verifier for which finding $t$ disjoint solutions is roughly as easy as finding a single solution. That said, the construction is somewhat lacking as both are able to find the desired set with relatively high probability, roughly $\frac{1}{2}$. As such, to prove \Cref{lem:lb}, we will take this construction and lift it to a harder instance via known direct product theorems. Indeed, we draw on the following result.

\begin{theorem}[Drucker \cite{drucker2012improved}]
\label{thm:dpt}
Let $\calX$ be a set, $\calR: \calX \rightarrow \zo$ be a random oracle, and $V^{\calR}$ be a verifier that makes queries to $\calR$. Suppose that no $Q$-query algorithm can produce an accepting solution to a verifier $V^{\calR}$ with probability more than $p$, where $p \leq 1 - 1/\ell$. Then if $\calR_1, \dots, \calR_\ell$ are independent random oracles, any algorithm making at most $Q$ queries in total across them queries verifying sets to $V^{\calR_1}, V^{\calR_2}, \dots V^{\calR_\ell}$ with probability at most $p^{\ell}$.
\end{theorem}

With this, we can then use the direct product of the instance of \Cref{lem:raw-lb} to get our proof of \Cref{lem:lb}.

\begin{proof}[Proof of \Cref{lem:lb}]
    Set $\ell = t^{7/\alpha}$ and consider independent oracles $\calR_1, \calR_2, \dots, \calR_\ell: \calX' \rightarrow \zo$ and let $V^{\calR_i}$ denote the verifiers as in \Cref{lem:raw-lb}. We will then consider the verifier $V^{\otimes \ell}(w_1, w_2, \dots, w_\ell)$ that accepts if and only if $V^{\calR_i}(w_i)$ accepts for all $i \in [\ell]$. Note that we can think of $V^{\otimes \ell}$ as querying a single random oracle over $\calX = \calX' \times [\ell]$. Note that with this setup, we have that $k = O(t^7 \ell) = \log^{O(1/\alpha)}(|\calX|)$ as each verifier makes $\textsf{poly}(t)$ queries and $|\calX'| = \exp(\textsf{poly}(t))$. We can similarly verify that $\ell \geq k^{1 - O(\alpha)}$.

    We now first note that there is an algorithm that outputs $t$ disjoint solutions with probability at least $2^{-\ell}$. Indeed, we can take the algorithm from \Cref{lem:raw-lb} and run it on each oracle independently. It succeeds with probability at least $2^{-\ell}$ and outputs $t$ disjoint solutions to $V^{\otimes \ell}$. Moreover, since the algorithm makes $\textsf{poly}(t)$ queries to each of the $\ell$ oracles, we then make $O(\ell \cdot \textsf{poly}(t)) = \textsf{poly}(k)$ queries in total, as desired. Thus, we get that $\eta_t \geq 2^{-\ell}$.

    On the other hand, we can apply the direct product theorem to bound the probability an algorithm finds a single solution. To do so, we first modify such an algorithm to query each solution it outputs. Notably, this adds at most $O(\ell t^7) = \textsf{poly}(k)$ queries to the algorithm. Given such an algorithm, applying \Cref{thm:dpt} then yields that no such algorithm making at most $2^{O(t)}$ queries can output a single solution with probability better than
        \[(1/2 + O(1/t))^\ell \leq 2^{-\ell} \cdot e^{O(\ell/t)} \leq 2^{-\ell/2}.\]
    From this we can note that any algorithm, which may or may not query what is sent to the verifier, that makes at most
        \[2^{O(t)} - \textsf{poly}(k) = 2^{k^{O(\alpha)}} - \textsf{poly}(k) = 2^{k^{O(\alpha)}} \]
    queries, succeeds with probability at most $2^{-\ell/2}$. Thus, $\eta_1 \leq 2^{-\ell/2}$.

    Putting the two pieces together, we then have
        \[ \eta_t \geq 2^{-\ell} \geq \eta_1^{t} \cdot 2^{t \ell/4} \geq \eta_1^t \cdot e^{t k^{1 - O(\alpha)}}. \]
\end{proof}

\subsection{Proof of \Cref{lem:raw-lb}}
At a high level, we will design our construction such that when a lucky event occurs we can find many disjoint verifying sets, yet when the event does not occur it is quite difficult to find a single solution. Roughly, we will do this by choosing a core of $N := t^8$ points $z_1, \dots, z_N$. We will aim to make it easy to produce many disjoint witnesses when $\calR(z) := \calR(z_1), \dots, \calR(z_N)$ has at most $N/2$ ones. Towards this, we will have the verifier take as input a string $x \in \zo^N$, which the reader should think of as a claimed value for $\calR(z)$. Crucially, the verifier will always reject any string $x$ with more than $N/2$ ones.

Now the verifier's goal is roughly to check that $x = \calR(z)$. That said, we need the verifier to query disjoint points in the oracle for each witness, so we instead will partition $[N]$ into $t$ pieces and have the algorithm submit an index $i$ to indicate what piece to check equality on. Notably, using a fixed partition $P_1, \dots, P_t$ is a poor choice as it will be easy to find a single verifying set by lying about the value of $x$. Namely, we can simply set $x_j = \calR(z_j)$ for $j \in P_i$ and $x_j = 0$ elsewhere.

To circumvent this, we will effectively assign a random partition to each $x \in \zo^N$. More explicitly, for each $x$, we associate a set of $t^2$ inputs $(x,y_1), \dots (x, y_{t^2})$. For each possible random string $r(x) := \calR((x,y_1)) \calR((x,y_2)) \dots \calR((x,y_{t^2}))$, we assign a partition of $[N]$ into $t$ sets, $P_1(r), \dots, P_t(r)$ for a mapping we'll give later. With this, the final verifier then takes as input $(x,s,i)$ with $x \in \zo^N$, $s \in \zo^{t^2}$, and $i \in [t]$. First and foremost, it rejects if $x$ has more $1$'s than $0$'s. Next, it rejects if $x_j \not = \calR(z_j)$ for some $j \in P_i(s)$. Finally, we must check that we used the correct partition. Towards this, we split the bits in $r(x) = \calR((x,y_1)) \calR((x,y_2)) \dots \calR((x,y_{t^2}))$ into $t$ blocks, i.e. $r|_{[t]}, r|_{(t,2t]}, \dots r|_{(t^2-t,t^2]}$. We then check that $s_j = r_j = \calR((x,y_j))$ for all indices $j$ in the $i$th block, i.e. for all $j \in (t(i-1), ti]$.

We can now quickly note that

\begin{lemma}
\label{lem:t-disj-raw}
    When $\calR(z)$ contains at most $N/2$ ones, there exists an algorithm that finds $t$ disjoint verifying sets to the verifier.
\end{lemma}

\begin{proof}
    We simply query the random oracle on $z_1, \dots, z_N$ and let $x = \calR(z)$. Note that by assumption $x$ has at most $N/2$ ones. We then query $\calR((x,y_1)) \calR((x,y_2)) \dots \calR((x,y_{t^2}))$ and set $s = r(x)$. Finally, we submit the witnesses $(x,s,i)$ for $i \in [t]$ as our disjoint witnesses. By construction, these witnesses never have overlapping queries. Indeed, $P_1(s), \dots, P_t(s)$ is a partition and each witness only checks one class. Similarly, each witness checks the random oracle on disjoint blocks of $(x,y_1), (x, y_2), \dots, (x,y_{t^2})$. Moreover, we can see that the verifier will accept each solution as $x$ has small enough Hamming weight and we chose $x$ and $s$ to agree with $\calR(z)$ and $r(x)$.
\end{proof}

We'll now argue that when $\calR(z)$ has large Hamming weight, which we will define to be at least $\frac{N}{2} + 4t^3$, then no algorithm that makes few queries can find a solution. To do this, we will need to choose our mapping of random strings to partitions. In particular, we choose any mapping from $\zo^{t^2}$ to any family of partitions that satisfies the below lemma.

\begin{lemma}
\label{lem:partitions}
    There exists a family $\calP$ of $2^{t^2}$ partitions of $[N]$ into $t$ equal-size pieces such that for any string $\Delta \in \zo^N$ with at least $4t^3$ non-zero entries
    \[ \left | \left \{ (P_1, \dots, P_t) \in \calP \bigg | \exists i \in [t]: \Delta|_{P_i} = 0 \right \} \right| \leq N  \]
\end{lemma}

Intuitively, the reader should think of $\Delta$ as corresponding to the entries where $x$ and $\calR(z)$ disagree. Notably, when $\calR(z)$ has high Hamming weight, it must disagree with any $x$ that has at most $N/2$ ones on at least $4t^3$ entries. Thus, the above lemma states that every piece of almost every partition will contain a point of disagreement between $x$ and $\calR(z)$.

\begin{proof}
    We prove the statement by the probabilistic method. Let $\calP$ be $2^{t^2}$ independent random equipartitions of $[N]$ into $t$ pieces. For a particular string $\Delta$ with at least $4t^3$ non-zero entries and a random partition $(P_1, \dots, P_t)$ we have that

        \[\Pr_{P_i}[\Delta|_{P_i} = 0] \leq \frac{\binom{N - 4t^3}{N/t}}{\binom{N}{N/t}} \leq \left( \frac{N-4t^3}{N} \right)^{N/t} \leq e^{-4t^2} \]

    We now take a union bound over all $i \in [t]$, to get that
        \[\Pr_{P_1, \dots, P_t}[\exists i \in [t]: \Delta|_{P_i} = 0] \leq t e^{-4t^2}. \]

    Thus, we can compute
    \begin{align*}
        \E_{\calP} \left[ \sum_{\Delta: |\Delta| \geq 4t^3} \mathbb{I} \left[ \left | \left \{ (P_1, \dots, P_t) \in \calP \bigg | \exists i \in [t]: \Delta|_{P_i} = 0 \right \} \right| > N \right] \right] &\leq 2^N \cdot 2^{Nt^2} \cdot \left( te^{-4t^2} \right)^N\\
        &< 1
    \end{align*}
    where we used the fact that there are at most $2^N$ values for $\Delta$ and at most $2^{Nt^2}$ ways to choose $N$ partitions from $\calP$. Thus, it must follow that there exists a $\calP$ satisfying the desired property.
\end{proof}

With this, we can now show that no algorithm making few queries succeeds at finding a verifying set when $\calR(z)$ has large Hamming weight.

\begin{lemma}
\label{lem:single-raw}
    Suppose that $t$ is sufficiently large and $\calR(z)$ has at least $\frac{N}{2} + 4t^3$ ones, then any algorithm that makes at most $2^{O(t)}$ queries to the random oracle finds a verifying set with probability at most $O(\frac{1}{t})$.
\end{lemma}

\begin{proof}
    We will throughout condition on the value of $\calR(z)$ and assume it is some fixed string with at least $N/2 + 4t^3$ ones. This suffices by noting that for any algorithm $\alg$
        \[\Pr \left[ \alg \text{ finds an accepting witness }\right] = \sum_{\sigma} \frac{1}{2^N} \cdot \Pr \left[ \alg \text{ finds an accepting witness } \bigg | \calR(z) = \sigma \right]. \]

    We will additionally assume that the algorithm runs the verifier before submitting its verifying set and thus queries everything the verifier does. In the end, this will increase the number of queries by at most $\frac{N}{t} + t$. With this, we will now bound the probability that an algorithm queries a verifying set. To do this, we consider all values of $x$ that the algorithm queries. That is, let $x^{(1)}, \dots, x^{(Q)}$ denote the distinct values in $\zo^N$, in order of first query, such that for each $a \in [Q]$ the algorithm queries some point $(x^{(a)},y_j)$ for some $j \in [t^2]$.
    For each $a \in [Q]$, we bound the probability that $x^{(a)}$ yields an accepting witness given that no earlier string does:
        \[\Pr_{\calR} \left [\exists s,i: V^{\calR}(x^{(a)},s,i) \text{ accepts} \bigg | V^{\calR}(x^{(b)},s,i) \text{ rejects} \quad \forall b < a, s \in \zo^{t^2}, i \in [t] \right] \]

    Note that to accept some witness $(x^{(a)},s,i)$ where $x^{(a)}$ has at most $N/2$ ones, we must have that $x^{(a)}$ and $\calR(z)$ agree on $P_i(s)$, so $P_1(s), \dots, P_t(s)$ is one of the at most $N$ bad partitions as defined by the condition in \Cref{lem:partitions}. As such, there are at most $N$ candidate values of $s$ that we must union bound over. For each one, we can easily verify that it agrees with $r(x^{(a)})$ on the $i$th contiguous block of the $t^2$ bits with probability $2^{-t}$, as must be the case in order to not be rejected when the verifier cross checks $s$ and $r(x^{(a)})$. Thus, taking a union bound over the $t$ values of $i$ and the at most $N$ bad partitions yields
    \[\Pr_{\calR} \left [\exists s,i: V^{\calR}(x^{(a)},s,i) \text{ accepts} \right] \leq N \cdot t \cdot 2^{-t}.\]

    Moreover, the above only relies on the randomness of $\calR((x^{(a)}, y_1)), \calR((x^{(a)}, y_2)), \dots, \calR((x^{(a)}, y_{t^2}))$. Since these values of the oracle are unqueried before $x^{(a)}$ is first queried and independent of those relevant for the previous rejections for $x^{(b)}$ with $b < a$, it follows that
        \[\Pr_{\calR} \left [\exists s,i: V^{\calR}(x^{(a)},s,i) \text{ accepts} \bigg | V^{\calR}(x^{(b)},s,i) \text{ rejects} \quad \forall b < a, s \in \zo^{t^2}, i \in [t] \right] \leq N \cdot t \cdot 2^{-t}.\]

    Thus, any such algorithm that makes at most $2^{t/2}$ queries, queries such a solution with probability at most
        \[1 - \left(1 - N \cdot t \cdot 2^{-t} \right)^{2^{t/2}} \leq 2^{-t/2} \cdot N \cdot t \leq \frac{1}{t} \]
    assuming $t$ is sufficiently large. Thus, for any algorithm that makes at most
        \[2^{t/2} - \frac{N}{t} - t \geq 2^{t/3}\]
    queries, it outputs, as opposed to having to query a solution, an accepting solution with probability at most $\frac{1}{t}$, as desired.
\end{proof}

We can now combine these pieces together to get the desired result.

\begin{proof}[Proof of \Cref{lem:raw-lb}]
    We start by noting that we have that $|\calX| = N + t^2 \cdot 2^N$, corresponding to $z_1, \dots, z_N$ and the $t^2$ random inputs for each $x \in \zo^N$. Moreover, the verifier makes $\frac{N}{t} + \frac{t^2}{t} = t^7 + t = O(t^7)$ queries as desired. Since $\calR(z)$ has at most $N/2$ ones with probability at least $\frac{1}{2}$, we have there exists an algorithm that finds $t$ disjoint solutions with probability at least $\frac{1}{2}$ using $\textsf{poly}(t)$ queries by \Cref{lem:t-disj-raw}, giving property $(i)$. Similarly, we can compute that $\calR(z)$ has fewer than $N/2 + 4t^3$ ones with probability at most $\frac{1}{2} + O(4t^3/\sqrt{N}) = \frac{1}{2} + O(1/t)$. Combining this with \Cref{lem:single-raw}, we then get that no algorithm making $2^{O(t)}$ queries can find a single solution with probability better than $\frac{1}{2} + O(1/t)$, yielding property $(ii)$.
\end{proof}

\paragraph{AI Usage.}

The authors initially proved a weaker version of the main result without AI assistance; the lower bound on the verifier complexity was essentially the same, provided that the commitment scheme had \emph{quasipolynomial security}. The authors determined that in order to extend this proof to the full case of polynomial security, it would suffice to prove the disjoint product lemma \footnote{Affectionately referred to by the authors as the Jonah Crab Conjecture (cf. \Cref{foot:crab})}. In the original proof, the authors established a weaker form of this lemma \footnote{Dubbed the Hermit Crab Lemma by the authors (cf. \Cref{foot:crab})} which was sufficient in the regime of quasi-polynomial query attackers (this was achieved by the coloring argument and direct product lemma discussed in the technical overview for intuition). Afterwards, the authors prompted ChatGPT Astra to prove the stronger disjoint product lemma. This proof was digested and rewritten by the authors and yields the main lower bound that appears in this work. Astra\footnote{This was prompted through the author's ChatGPT dot.} also provided a proof of the barrier to stronger forms of the disjoint product lemma appearing in \Cref{lem:lb-intro}, which the authors iterated with the models to simplify, digested, and rewrote here. Outside of the disjoint product lemma and this barrier, the current proof is essentially the same as that originally conceived by the authors prior to interaction with Astra; machine assistance was used in many places to reorganize and simplify the proof and to tighten various parameters.
\Cref{fig:merkle-binary,fig:merkle-dary} were made using interaction with Claude Opus.

 \paragraph{Acknowledgments}

The authors thank Yuval Efron and Sam Gunn for numerous valuable discussions and for participating in earlier stages of this work.
Miranda Christ is supported by an Amazon Research Award, Google CyberNYC, NSF CCF-2312224, NSF CCF-2107187, NSF CCF-2332922, the UK AISI Alignment Project, and The Leighton Family Chair.
Oliver Korten is supported by the National Science Foundation under Grant No.~DMS-2424441.
Tal Malkin is supported by a Google CyberNYC grant, an Amazon Research Award, and NSF grant CCF-2312242.
Noam Mazor is supported by the Simons Foundation.
Shyamal Patel is supported by the NSF AI Institute for Foundations of Machine Learning (IFML), NSF grants CCF-2106429, CCF-2107187, CCF-2218677, ONR grant ONR-13533312, and an NSF Graduate Student Fellowship.

\newpage

\bibliographystyle{alpha}
\bibliography{refs}

\end{document}